\documentclass[12pt,preprint]{aastex}
\usepackage{natbib}
\usepackage{color}
\usepackage{graphicx}
\usepackage{rotating}
\usepackage{hyperref}
\usepackage[all]{hypcap}
\usepackage{mathrsfs}
\hypersetup{
    bookmarks=true,         
    unicode=false,          
    pdftoolbar=true,        
    pdfmenubar=true,        
    pdffitwindow=false,     
    pdfstartview={FitH},    
    pdftitle={My title},    
    pdfauthor={Author},     
    pdfsubject={Subject},   
    pdfcreator={Creator},   
    pdfproducer={Producer}, 
    pdfkeywords={keyword1} {key2} {key3}, 
    pdfnewwindow=true,      
    colorlinks=true,        
    linkcolor=blue,         
    citecolor=blue,         
    filecolor=magenta,      
    urlcolor=cyan           
}

\begin{document}
\title{Magnetic Helicity Estimations in Models and Observations of the Solar Magnetic Field. Part III: Twist Number Method}
\author{Y. Guo\altaffilmark{1}, E. Pariat\altaffilmark{2}, G. Valori\altaffilmark{3}, S. Anfinogentov\altaffilmark{4}, F. Chen\altaffilmark{5}, M. Georgoulis\altaffilmark{6}, Y. Liu\altaffilmark{7}, K. Moraitis\altaffilmark{2,6}, J. K. Thalmann\altaffilmark{8}, S. Yang\altaffilmark{9}}

\affil{$^1$ School of Astronomy and Space Science and Key Laboratory of Modern Astronomy and Astrophysics in Ministry of Education, Nanjing University, Nanjing 210023, China} \email{guoyang@nju.edu.cn}
\affil{$^2$ LESIA, Observatoire de Paris, PSL Research University, CNRS, Sorbonne Universit\'e, UPMC Univ. Paris 06, Univ. Paris Diderot, Sorbonne Paris Cit\'e, 92190 Meudon, France}
\affil{$^3$ University College London, Mullard Space Science Laboratory, Holmbury St. Mary, Dorking, Surrey, RH5 6NT, United Kingdom}
\affil{$^4$ Institute of Solar-Terrestrial Physics SB RAS 664033, Irkutsk, PO box 291, Lermontov street, 126a, Russia}
\affil{$^5$ Max-Plank-Institut f\"ur Sonnensystemforschung, 37077 G\"ottingen, Germany}
\affil{$^6$ Research Center for Astronomy and Applied Mathematics of the Academy of Athens, 4 Soranou Efesiou Street, 11527 Athens, Greece}
\affil{$^7$ W. W. Hansen Experimental Physics Laboratory, Stanford University, Stanford, CA 94305-4085, USA}
\affil{$^8$ Institute of Physics, Univeristy of Graz, Universit\"atsplatz 5/II, 8010 Graz, Austria}
\affil{$^{9}$ Key Laboratory of Solar Activity, National Astronomical Observatories, Chinese Academy of Sciences, Beijing 100012, China}

\begin{abstract}
We study the writhe, twist and magnetic helicity of different magnetic flux ropes, based on models of the solar coronal magnetic field structure. These include an analytical force-free Titov--D\'emoulin equilibrium solution, non force-free magnetohydrodynamic simulations, and nonlinear force-free magnetic field models. The geometrical boundary of the magnetic flux rope is determined by the quasi-separatrix layer and the bottom surface, and the axis curve of the flux rope is determined by its overall orientation. The twist is computed by the Berger--Prior formula that is suitable for arbitrary geometry and both force-free and non-force-free models. The magnetic helicity is estimated by the twist multiplied by the square of the axial magnetic flux. We compare the obtained values with those derived by a finite volume helicity estimation method. We find that the magnetic helicity obtained with the twist method agrees with the helicity carried by the purely current-carrying part of the field within uncertainties for most test cases. It is also found that the current-carrying part of the model field is relatively significant at the very location of the magnetic flux rope. This qualitatively explains the agreement between the magnetic helicity computed by the twist method and the helicity contributed purely by the current-carrying magnetic field.
\end{abstract}


\section{Introduction}
A magnetic flux rope is a key ingredient for various solar-activity models, such as filament/prominence eruptions, flares, and coronal mass ejections (CMEs). The terms magnetic flux rope and magnetic flux tube are defined as a bundle of magnetic field lines of finite size twisting around a common axis curve. When a flux tube is infinitesimally thin, it represents a single magnetic field line. To study the equilibrium and stability of a magnetic flux rope, it is crucial to know its force balance, free magnetic energy, and magnetic helicity. There are various ways to quantify the force and energy of a magnetic flux rope based on theoretical, numerical, and observational methods \citep[e.g.][]{1996Chen,1998Lin,2002Regnier,2005Torok,2006Kliem,2013Feng}. The method of choice to quantify the magnetic helicity of a flux rope is still an open issue, because there are various uncertainties in observations, models, and methods. These include, e.g., limitations of current techniques to measure and model the magnetic field in the solar atmosphere, difficulties in quantifying the topological boundaries of flux ropes, as well as the uncertainty of helicity computations.

Magnetic helicity quantitatively measures the geometrical complexity of a magnetic field. A gauge invariant helicity measure for open magnetic configurations (with field lines penetrating the boundaries) is defined by the relative magnetic helicity in finite volumes \citep{1984Berger,1985Finn}:
\begin{equation}
\mathscr{H}_V = \int_V(\mathbf{A}+\mathbf{A}_\mathrm{p}) \cdot (\mathbf{B} - \mathbf{B}_\mathrm{p}) \mathrm{d}V , \label{eqn:hv}
\end{equation}
where $\mathbf{B}$ is the vector magnetic field in volume $V$, $\mathbf{B}_\mathrm{p}$ is the reference magnetic field that is usually selected as the potential field with the same normal magnetic field as $\mathbf{B}$ on the boundaries of $V$, $\mathbf{A}$ is the vector potential of $\mathbf{B}$ with $\mathbf{B} = \nabla \times \mathbf{A}$, and $\mathbf{A}_\mathrm{p}$ is the vector potential of the potential field $\mathbf{B}_\mathrm{p}$. The relative magnetic helicity is a global quantity. Its local density in an arbitrary volume does not have a physical meaning, because the vector potential depends on the distribution of the field in the entire volume, and because adding a gauge function to any vector potential would arbitrarily change the local helicity-density values. However, magnetic helicity does have a local density per elementary flux tube, namely, the field line helicity defined as the integral of $\mathbf{A}$ along a magnetic field line \citep{2014Yeates,2016Yeates,2015Russell}. Besides the relative magnetic helicity in Equation~(\ref{eqn:hv}), there are some other expressions and interpretations of the magnetic helicity \citep{1984Jensen,2006Hornig,2006Low,2011Low,2008Longcope,2014Prior}. Here, we only focus on the relative magnetic helicity.

There are several practical ways to compute magnetic helicity either using a finite volume method \citep{2011Rudenko,2011Thalmann,2012Valori,2013Yang,2014Rudenko,2014Moraitis}, a twist number method \citep{2010Guo,2013Guo}, a helicity-flux integration method \citep{2001Chae,2005Pariat,2012Liu}, or a connectivity-based method \citep{2012Georgoulis}. These methods differ in their input magnetic field and in the way to calculate the magnetic helicity. The finite volume method employs Equation~(\ref{eqn:hv}) and requires the full 3D magnetic field vector information as an input. The twist number method also requires the magnetic field in a 3D volume but with the additional requirement that a magnetic flux rope is present. The helicity-flux integration method requires a time series of two-dimensional (2D) vector magnetic field and velocity maps as an input. Consequently, with this method, only the accumulation of the magnetic helicity injected through a 2D surface can be computed. The connectivity-based method requires only a single vector magnetic field map on the bottom boundary and assumes that the magnetic polarities are magnetically connected over a minimal connection length. A detailed description of all of these methods is presented in the first paper of a series \citep{2016Valori}, where different implementations of the finite volume method are also compared. The comparison of existing implementations of the flux integration method and the connectivity-based method is the subject of a second paper \citep{2017Pariat}, while a third paper of the series will implement different helicity methods on a particularly observed eruptive solar active region (Georgoulis et al. 2017, in preparation).

The twist number method estimates the magnetic helicity of a magnetic flux rope by computing the twist and axial magnetic flux. The twist measures the rotation of an individual field line about the central axis of the flux rope (i.e., the axis curve). Figures illustrating how the twist is measured can be found in Figure~1 of \citet{2006Berger} and Figure~3 of \citet{2012Prior}. When computing the helicity of isolated flux rope structures, one basically ignores the magnetic field surrounding the structure, and its connection to the field inside of the rope. However, it is still meaningful to compute the magnetic helicity of a magnetic flux rope, for three reasons: first, the solar active-region corona could be approximated by a major electric current channel embedded in a potential field using the argument of \citet{1999Titov}. Second, the lateral boundary of a magnetic flux rope is a magnetic flux surface without any magnetic flux penetrating it; therefore, the magnetic helicity within the magnetic flux surface is conserved under an ideal evolution. Third, \citet{2003Berger} showed that $\mathscr{H}_V$ can be decomposed into two separately gauge invariant components:
\begin{equation}
\mathscr{H}_V = \mathscr{H}_{V,J} + \mathscr{H}_{V,JP}, \label{eqn:hv2}
\end{equation}
with
\begin{eqnarray}
\mathscr{H}_{V,J} & = & \int_V (\mathbf{A} - \mathbf{A}_\mathrm{p}) \cdot (\mathbf{B} - \mathbf{B}_\mathrm{p}) \mathrm{d}V  , \label{eqn:hvj} \\
\mathscr{H}_{V,JP} & = & 2\int_V \mathbf{A}_\mathrm{p} \cdot (\mathbf{B} - \mathbf{B}_\mathrm{p}) \mathrm{d}V . \label{eqn:hvjp}
\end{eqnarray}
$\mathscr{H}_{V,J}$ measures the magnetic helicity contributed purely by the magnetic field that carries local currents. The other part $\mathscr{H}_{V,JP}$ is the mixed contribution of the magnetic helicity between the potential magnetic field and the magnetic field generated by local currents. The magnetic helicity of a magnetic flux rope would have a physical meaning if it contributed a major part of $\mathscr{H}_{V,J}$ in the entire volume $V$.

To guarantee the gauge invariance of the magnetic helicity, the input magnetic field should be solenoidal. To quantify the solenoidal condition of a magnetic field, \citet{2013Valori} proposed to use Thomson's theorem, which decomposes the magnetic field into four parts, namely, the solenoidal potential, solenoidal current-carrying, non-solenoidal potential, and non-solenoidal current-carrying parts. Correspondingly, the associated magnetic energy can also be decomposed into four terms, plus a fifth term accounting for mixed contributions. If the total magnetic energy, including all of the aforementioned terms, is denoted by $E$, and the total non-solenoidal energy (potential, current-carrying, mixed) is denoted by $E_\mathrm{ns}$ (also refer to Appendix A of \citealt{2016Valori}), the ratio between $E_\mathrm{ns}$ and $E$ can be used to quantify how well the solenoidal condition is fulfilled. The numbers listed in Table~1 of \citet{2013Valori} provide a quantitative comparison between $E_\mathrm{ns}/E$ and $<|f_i|>$, which is the average of the absolute value of the fractional flux change in a numerical cell (refer to Appendix C of \citealt{2013Valori}). We note that $<|f_i|>$ is another size-dependent measure of the solenoidality of the field. For the test cases considered in \citet{2013Valori}, if $E_\mathrm{ns}/E$ is less than $2\%$, $<|f_i|>$ is less than $2 \times 10^{-5}$ (see, e.g., the first three rows of their Table 1). An alternative method to assess the non-solenoidality of the field via the magnetic energy is described in \citet{2014Moraitis}.

In \citet{2016Valori}, a preliminary comparison of the twist number method and the finite volume method is presented. It was found that the magnetic helicity estimated by the twist number method approximately matches the purely current-carrying part, namely $\mathscr{H}_{V,J}$, of the total relative magnetic helicity of the Titov--D\'emoulin model. It was also shown that the magnetic helicity estimated by the twist number method matches $\mathscr{H}_{V,J}$ better for higher twist and spatial resolution. In order to progress in the quantification of the abilities of the twist number method, here we will test how the twist number depends on the choice of the axis of a magnetic flux rope in the Titov--D\'emoulin model. We provide a systematic study on the performance of the twist number method when applied to various magnetic field models and compare it to the results delivered from a finite volume method. We also provide an analysis of the magnetic fields by splitting them into a potential part and a current-carrying part, in order to explain why the magnetic helicity estimated by the twist number method matches $\mathscr{H}_{V,J}$ rather than $\mathscr{H}_V$.

The outline of the paper is as follows. The twist number method is described is Section~\ref{sec:method}. Results of the method's application to the Titov--D\'emoulin model \citep{1999Titov}, magnetohydrodynamic (MHD) numerical simulations \citep{2013Leake,2014Leake}, and nonlinear force-free field (NLFFF) models \citep{2015Savcheva,2016Savcheva} are presented in Section~\ref{sec:result}. We finally provide a summary and make a discussion in Section~\ref{sec:discussion}.

\section{The Twist Number Method} \label{sec:method}
With the Titov--D\'emoulin model, MHD numerical simulations, and NLFFF models computed by the flux rope insertion method, one can obtain 3D magnetic field models hosting magnetic flux ropes. In order to quantify the magnetic helicity of these flux ropes, one needs to determine their geometrical boundaries. The quasi-separatrix layer \citep[QSL;][]{1995Priest,1996Demoulin,1997Demoulin,2002Titov} is a useful concept to serve such a purpose. QSLs are 3D thin volumes where the gradient of the magnetic field line connectivity is large, as measured by the squashing degree $Q$ \citep{1996Demoulin,2002Titov,2007Titov}. \citet{2012Pariat} compared three different methods and identified a best-performing method to compute the squashing degree $Q$ in a 3D volume. This method has been implemented by various authors and applied to analyze the magnetic topology of magnetic fields derived by NLFFF extrapolations \citep{2014Zhao,2015Yang,2016Yang,2016Liu}. It has been shown that magnetic flux ropes are associated with bald patches or hyperbolic flux tubes \citep{1999Titov,2002Titov}. Bald patches are locations along the polarity inversion line where the field is shaped concave-up and oriented tangent to the photosphere, whereas hyperbolic flux tubes are volumes defined by the intersection of two or more QSLs. The QSLs associated with these topology structures wrap the magnetic flux ropes and separate them from their surroundings. \citet{2013Guo} also found that a magnetic flux rope wrapped by QSLs, based on an NLFFF model using the optimization algorithm of \citet{2004Wiegelmann}.

\citet{1984Berger} have assigned the magnetic helicity a clear geometrical meaning. They pointed out that magnetic helicity quantitatively measures the geometrical complexity of magnetic field lines. If a magnetic configuration consists of a finite number, $N$, of flux tubes, the magnetic helicity is determined by the linkage and knotting of different flux tubes (mutual helicity), and by the writhe and twist of all of the individual flux tubes (self helicity). For closed curves, the linking number, twist, and writhe are well defined as shown in \citet{2006Berger}. The Gauss linking number measures the mutual linkage of two curves $\mathbf{x}(s)$ and $\mathbf{y}(s')$ that are parameterized by $s$ and $s'$:
\begin{equation}
\mathcal{L} = \frac{1}{4\pi} \oint_{\mathbf{x}} \oint_{\mathbf{y}} \mathbf{T}_\mathbf{x}(s) \times \mathbf{T}_\mathbf{y}(s') \cdot \frac{\mathbf{r}}{|\mathbf{r}|^3} \mathrm{d}s' \, \mathrm{d}s , \label{eqn:linking}
\end{equation}
where $\mathbf{T}_\mathbf{x}(s)$ and $\mathbf{T}_\mathbf{y}(s')$ are the unit tangent vector to $\mathbf{x}(s)$ and $\mathbf{y}(s')$, respectively, and $\mathbf{r}$ is the position vector with $\mathbf{r}=\mathbf{x}(s) - \mathbf{y}(s')$. The writhe measures the non-planarity of a single curve:
\begin{equation}
\mathcal{W} = \frac{1}{4\pi} \oint_{\mathbf{x}} \oint_{\mathbf{x}} \mathbf{T}(s) \times \mathbf{T}(s') \cdot \frac{\mathbf{r}}{|\mathbf{r}|^3} \mathrm{d}s' \, \mathrm{d}s , \label{eqn:writhe}
\end{equation}
where the position vector $\mathbf{r}$ points from $\mathbf{x}(s')$ to $\mathbf{x}(s)$ such that $\mathbf{r}=\mathbf{x}(s) - \mathbf{x}(s')$. The twist measures the rotation amount of one curve $\mathbf{y}(s')$ about the other $\mathbf{x}(s)$:
\begin{equation}
\mathcal{T} = \frac{1}{2\pi} \oint_{\mathbf{x}} \mathbf{T}(s)\cdot \mathbf{V}(s) \times \frac{\mathrm{d}\mathbf{V}(s)}{\mathrm{d}s} \mathrm{d}s , \label{eqn:twist}
\end{equation}
where $\mathbf{V}(s)$ is a unit vector normal to $\mathbf{T}(s)$ and pointing from $\mathbf{x}(s)$ to $\mathbf{y}(s')$. The linking number and writhe are global quantities which involve double integrals of geometrical parameters along the curves. The linking number of two closed curves is always an integer. The linking number of a tube or ribbon (the surface between two non-intersecting curves forms a ribbon) equals the sum of the twist and writhe as demonstrated by the C$\mathrm{\check{a}}$lug$\mathrm{\check{a}}$reanu theorem \citep[e.g.,][]{1978Fuller,1992Moffatt,2006Berger}:
\begin{equation}
\mathcal{L} = \mathcal{W} + \mathcal{T}. \label{eqn:calug}
\end{equation} 

Magnetic field lines can be regarded as infinitesimally thin flux tubes and represented by curves in 3D space. In a closed configuration where no magnetic flux penetrates the boundaries, the total magnetic helicity is quantitatively expressed as the sum of the self and mutual helicity contributed by the $N$ flux tubes \citep{1984Berger,2006Demoulin}:
\begin{equation}
\mathscr{H} \approx \sum_{i=1}^N \mathcal{L}_i \Phi_i^2 + \sum_{i=1}^N \sum_{j=1,j \ne i}^N \mathcal{L}_{i,j} \Phi_i \Phi_j \, , \label{eqn:hel}
\end{equation}
where $\mathcal{L}_i$ denotes the sum of the twist and writhe of flux tube $i$ with magnetic flux $\Phi_i$, namely, $\mathcal{L}_i = \mathcal{W}_i + \mathcal{T}_i$, and $\mathcal{L}_{i,j}$ denotes the linking number of two flux tubes $i$ and $j$, respectively.

For open configurations, where flux penetrates the volume's surface, Equation~(\ref{eqn:hel}) still holds, only its meaning is changed. The magnetic helicity of open configurations is gauge invariant and physically meaningful in context with the definition of a relative magnetic helicity. To be applicable also to open curves, \citet{2006Demoulin} proposed an alternative definition of the linking number $\mathcal{L}_{i,j}$ following the concept of helicity injection. \citet{2006Berger} also proposed an alternative definition of the writhe for open curves. They used a directional expression, e.g., along the vertical direction $\mathbf{z}$, for the writhe to define a polar writhe $\mathcal{W}_p$. The key idea is that a curve is split into $n$ pieces by its local turning points where $\mathrm{d}s/\mathrm{d}z = 0$, and the double integration in Equation~(\ref{eqn:writhe}) can be expressed as the sum of a single integration. The polar writhe includes a local part, $\mathcal{W}_{pl}$, and nonlocal part, $\mathcal{W}_{pnl}$ \citep[see also,][]{2016Prior}:
\begin{equation}
\mathcal{W}_p = \mathcal{W}_{pl} + \mathcal{W}_{pnl}, \label{eqn:polar}
\end{equation}
\begin{equation}
\mathcal{W}_{pl} = \sum\limits_{i=1}^{n} \frac{1}{2\pi} \int_{z_i^\mathrm{min}}^{z_i^\mathrm{max}} \frac{\mathbf{z} \cdot \mathbf{T}_i \times \frac{\mathrm{d} \mathbf{T}_i}{\mathrm{d}z}}{1+|\mathbf{z} \cdot \mathbf{T}_i|} \mathrm{d} z, \label{eqn:local}
\end{equation}
\begin{equation}
\mathcal{W}_{pnl} = \mathop{\sum\limits_{i=1}^{n} \sum\limits_{j=1}^{n}}\limits_{i \ne j} \frac{\sigma_i \sigma_j}{2\pi} \int_{z_{ij}^\mathrm{min}}^{z_{ij}^\mathrm{max}} \frac{\mathrm{d}\Theta_{ij}}{\mathrm{d}z} \mathrm{d} z, \label{eqn:nonlocal}
\end{equation}
where $\sigma_i$ indicates whether the $i$-th piece of the curve exists at height $z$, and whether it is rising or falling. If $z \in (z_i,z_{i+1})$ and $\mathrm{d}s/\mathrm{d}z > 0$, $\sigma_i=1$; if $z \in (z_i,z_{i+1})$ and $\mathrm{d}s/\mathrm{d}z < 0$, $\sigma_i=-1$; if $z \notin (z_i,z_{i+1})$, $\sigma_i=0$. And $\Theta_{ij}$ is the azimuth angle of the position vector pointing from $\mathbf{x}_i(z)$ to $\mathbf{x}_j(z)$. Equations~(\ref{eqn:polar}), (\ref{eqn:local}), and (\ref{eqn:nonlocal}) have been adopted to compute the writhe of open curves, such as the helical structures in the corona \citep{2010Torok,2012Prior}. 

Since the twist is a local quantity with a well defined twist density:
\begin{equation}
\frac{\mathrm{d} \mathcal{T}}{\mathrm{d} s} = \frac{1}{2 \pi} \mathbf{T} \cdot \mathbf{V} \times \frac{\mathrm{d} \mathbf{V}}{\mathrm{d} s} , \label{eqn:twist2}
\end{equation}
the formula for closed curves is still applicable for open curves. Equation~(\ref{eqn:twist}) is the integration of Equation~(\ref{eqn:twist2}) along an axis curve. It is suitable for smooth curves in arbitrary geometries without self intersection. Therefore, it is also suitable for both force-free and non-force-free models. Equation~(\ref{eqn:twist}) has been applied to compute the twist in \citet{2010Guo,2013Guo}, \citet{2014Xia}, and \citet{2016Yang}. 

The twist number method is designed to estimate the magnetic helicity of a single highly twisted magnetic flux rope. Two major approximations are adopted for this method. On the one hand, the highly twisted magnetic structure is considered as a single flux tube, thus the mutual helicity between the flux tube and the surrounding magnetic field is omitted. This approach is motivated by observations of solar eruptions, since usually only one major magnetic flux rope is present in an active region. On the other hand, the magnetic helicity contributed by the writhe is also omitted. Observations and models of magnetic flux rope structures and evolutions suggest that they are usually not highly kinked due to low twists of magnetic field lines. The kink instability is not triggered in these cases. With the above two approximations, the magnetic helicity can be approximated by the twist of a single highly twisted structure as:
\begin{equation}
\mathscr{H}_\mathrm{twist} \approx \mathcal{T} \Phi^2 \, , \label{eqn:hel2}
\end{equation}
where $\mathcal{T}$ and $\Phi$ are the twist number and magnetic flux of the single magnetic flux rope. For the cases possessing highly kinked magnetic flux ropes \citep[e.g.,][]{2005Torok,2010Guo}, the magnetic helicity contributed by the writhe cannot be omitted, and the twist number method is not applicable to those structures that have significant writhe. But for the cases considered below, we will show that the writhe is small compared to the twist.

\section{Results} \label{sec:result}
The twist number method is applied to three different magnetic field models, all of them enclosing a magnetic flux rope. These are the Titov--D\'emoulin model \citep{1999Titov} in Section~\ref{sec:td}, MHD numerical simulations \citep{2013Leake,2014Leake} in Section~\ref{sec:mhd}, and NLFFF models \citep{2015Savcheva,2016Savcheva} in Section~\ref{sec:nlfff}. These models provide different challenges for the magnetic helicity estimation method. The Titov--D\'emoulin model is semi-analytically computed and serves a static 3D magnetic field solution, within which the existing flux rope is easily determined. The MHD simulations are time dependent while the NLFFF models are also static. But both are computed numerically and the flux rope structures are more complex than that of the Titov--D\'emoulin model. For example, the flux rope in the MHD simulations are more extended than that in the Titov--D\'emoulin model, and possess a lower twist, causing larger uncertainties in the computed twist helicity values. Similarly to the MHD simulations, the flux ropes in the NLFFF models have more complicated boundaries than that of the Titov--D\'emoulin model, also increasing the uncertainty in the computation of the magnetic helicity.

Some tests of the twist number method on the Titov--D\'emoulin model have already been performed and presented in Section 9.1 of \citet{2016Valori}. In this paper, we provide additional results for one case of the Titov--D\'emoulin model to provide more detailed information on the twist number method, such as the 3D QSL structure associated with the magnetic flux rope and the dependence of the twist on the position of the axis. We also provide a systematic analysis of the magnetic helicity computed by the twist number method for MHD and NLFFF models and compare the results with the finite volume method.

\subsection{Titov--D\'emoulin Model} \label{sec:td}

\begin{figure}
\begin{center}
\includegraphics[width=0.8\textwidth]{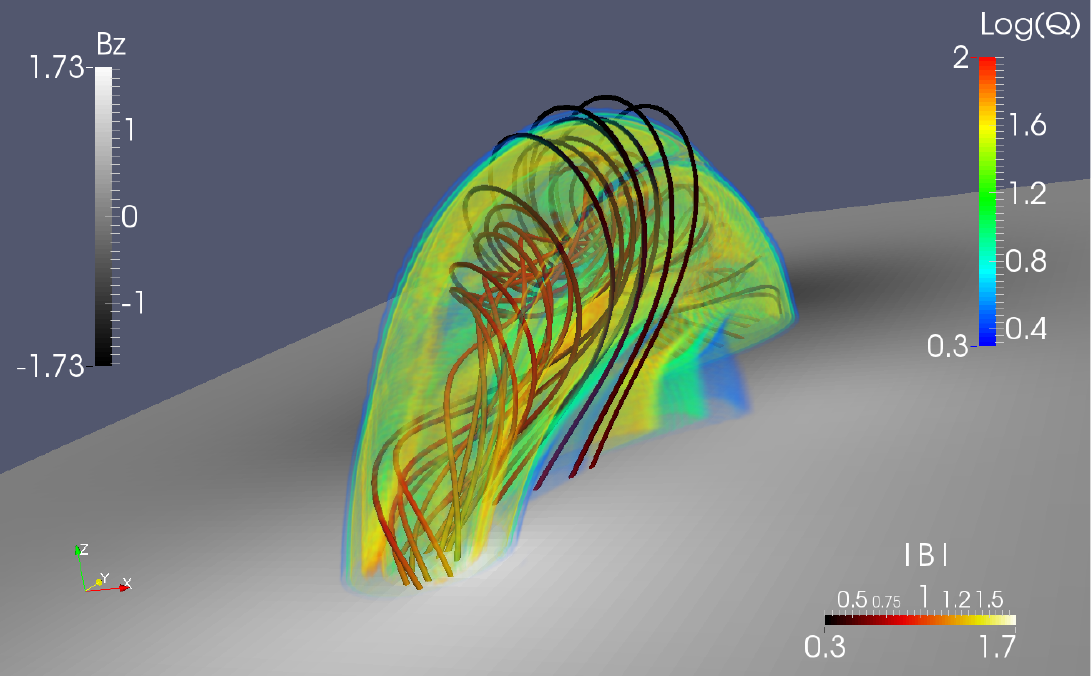}
\caption{QSLs and magnetic field lines in the magnetic flux rope of the Titov--D\'emoulin model, adhering to a twist of about three turns. Semi-transparent layers surrounding the magnetic field lines display the QSLs. Gray-scale images on the bottom show the vertical magnetic field $B_z$. The magnetic field lines are colored by the magnetic field strength $|B|$.
} \label{fig1}
\end{center}
(A movie attached to this figure is available online.)
\end{figure}

The dataset of the Titov--D\'emoulin model is similar to the one used in \citet{2016Valori}. The reader is referred to that study for further details on the magnetic field data. To derive the twist of a magnetic flux rope, we need its geometrical information, i.e., its boundary and axis curve. A magnetic flux rope is usually surrounded by a QSL since it is usually associated with bald patches or hyperbolic flux tubes, where the connections of magnetic field lines change rapidly \citep[e.g.,][]{1999Titov,2013Guo,2016Liu,2016Zhao}. For the magnetic flux rope as shown in Figure~\ref{fig1}, its boundary is determined by the bottom boundary and the QSLs. We adopt the method proposed by \citet{2012Pariat} to compute the 3D distribution of the squashing degree \citep{2002Titov,2007Titov}, $Q$, where large $Q$ values (with $Q \gg 2$) indicate the location of QSLs. Figure~\ref{fig1} and the supplementary movie display the QSLs and some magnetic field lines in the magnetic flux rope. The model configuration used here is identical with the ``TD-N3-0.06'' case in \citet{2016Valori}, with the model flux rope possessing a twist of about three full turns around its axis. It is clear to see that QSLs surround some magnetic field lines, which are regarded as the constituent part of the magnetic flux rope.

\begin{figure}
\begin{center}
\includegraphics[width=0.8\textwidth]{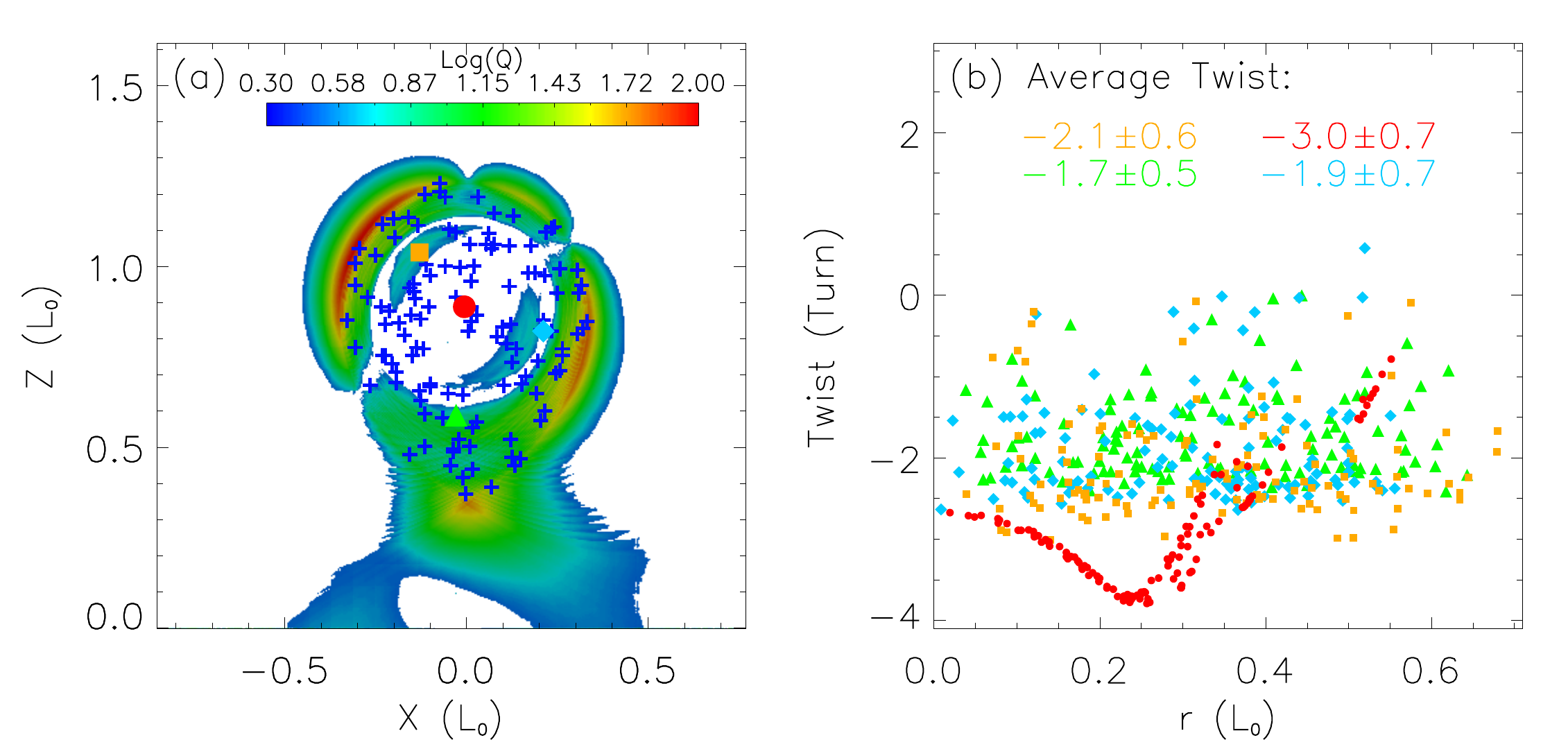}
\caption{(a) Vertical slice of the $Q$ map in the $xz$-plane at $y=0$, namely, in the middle of the flux rope along its axis. Red dot, orange square, cyan diamond, and green triangle signs indicate the intersections of the axis curves with the $xz$-plane for different test setups, in order to compare the effect of their relative locations on the retrieved flux rope twist. Blue plus sign indicates the intersections of the sample field lines, used to calculated the average flux rope twist, with the $xz$-plane. (b) Twist of the sample magnetic field lines along the distance, $r$ from the flux-rope axis. Symbols with different colors and signs represent the twist distribution for varying positions of the axis (shown in panel (a), using the same color and sign notation).
} \label{fig2}
\end{center}
\end{figure}

In order to determine the axis of the model flux rope, we make advantage of the symmetrical properties of its geometry. We assume the axis of the magnetic flux rope to be aligned with the $y$-axis, i.e., lying on a plane and thus possessing zero writhe. We first consider $Q$ within a vertical slice lying on the $xz$-plane at $y=0$. We delineate the projected boundary of the flux rope, based on the largest values of $Q$ (Figure~\ref{fig2}(a)) and assume all points within this boundary to be part of the flux rope. The latter can be used to compute the axial magnetic flux of the flux rope, in the form $\Phi = \int \int B_y \mathrm{d}x \mathrm{d}z$. If we assume that the magnetic field and the length are normalized by $B_0$ and $L_0$ in the Titov--D\'emoulin model, respectively, the magnetic flux of the flux rope is $\Phi = 0.17 B_0 L_0^2$. Then, the axis of the flux rope is determined as the magnetic field line with the minimum value of $B_r/|B_y|$, where $B_r = \sqrt{B_x^2 + B_z^2}$, oriented almost perpendicular to the $xz$-plane (red dot in Figure~\ref{fig2}(a)).

The twist of each sample field line (their cross sections with the vertical slice are indicated by blue crosses in Figure~\ref{fig2}(a)) as a function of distance from the axis curve of the flux rope is displayed in Figure~\ref{fig2}(b), represented by red dots. It shows that the twist first increases, until $r \approx 0.24 L_0$, and decreases for locations further away from the axis curve. Though the sample field lines obviously adhere to a different amount of twist, in order to use Equation~(\ref{eqn:hel2}), we aim to find a single number which quantifies the overall twist of the magnetic flux rope, where we use the arithmetic average of the field line's twist number. Furthermore, the standard deviation of the twists of all sample field lines quantifies the spread of the twists around their average number, thus can be regarded as an uncertainty measure. We find an average twist of $-3.0\pm0.7$ turns, and using Equation~(\ref{eqn:hel2}), a helicity of $\mathscr{H}_\mathrm{twist} = (-0.087 \pm 0.020) B_0^2 L_0^4$. The writhe of the axis curve is computed based on Equations~(\ref{eqn:polar}), (\ref{eqn:local}), and (\ref{eqn:nonlocal}) using a code available online\footnote{\url{https://www.maths.dur.ac.uk/~ktch24/code.html}}. We find $\mathcal{W}_p = -8.6\times 10^{-4}$, which is very small compared to the twist of the magnetic flux rope. It furthermore justifies the geometrical method used to determine the axis as perpendicular to the $xz$-plane.

In order to test the influence of the location of the axis curve of the flux rope on the result, we redefine its location of the intersection with the $xz$-plane at three different positions (other than that marked by the red dot in Figure~\ref{fig2}(a)). The twist numbers in Figure~\ref{fig2}(b) show that the average twists with axes at arbitrarily selected positions are smaller compared to the firstly analyzed situation, where the axis possesses the minimal poloidal magnetic flux. Furthermore, in the cases with displaced axis curves, the distributions of the twist numbers are much less coherent that the latter. This highlights how important the precise determination of the axis position is for a reliable estimation of its twist. In the TD-N3-0.06 case, when the axis is defined at the symmetrical position as indicated by the red dot in Figure~\ref{fig2}(a), the magnetic helicity computed by the twist method is closest to that derived by the finite volume method, which is $-0.090 B_0^2 L_0^4$ as listed in Table~6 of \citet{2016Valori}.

\subsection{MHD Numerical Simulated Models} \label{sec:mhd}

Two MHD models constructed by \citet{2013Leake} and \citet{2014Leake} are adopted here to extract the 3D magnetic field for the computation of magnetic helicity. The MHD models use the visco-resistive MHD equations to simulate a magnetic flux rope emergence from the upper convection zone into the corona. The two models differ in the strength and orientation of the overlying dipolar coronal magnetic field, which results in a stable (i.e., non-eruptive, named JL stable case hereafter) and unstable (i.e., eruptive, named JL unstable case hereafter) configuration for each case. From the original simulation datasets, only the coronal domain is extracted and the magnetic field is interpolated onto a uniform grid (for details see \citealt{2016Valori}). We select the following snapshots for the computation of magnetic helicity, namely, $t/t_0 = 30, 50, 85, 120, 155, 190$ for the stable case, and $t/t_0 = 30, 50, 80, 110, 140, 150$ for the unstable case, where $t_0$ is the normalization factor for time.

\begin{figure}
\begin{center}
\includegraphics[width=0.8\textwidth]{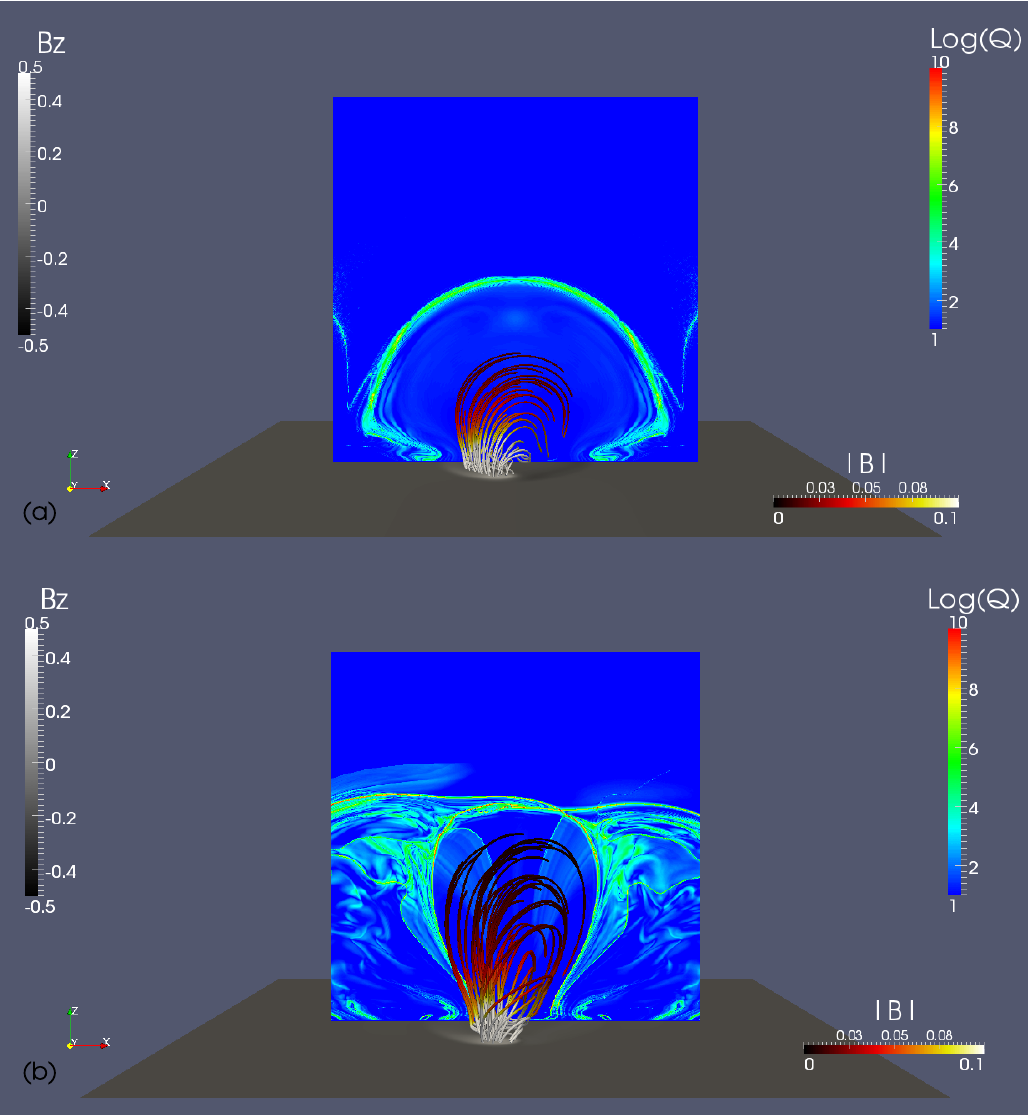}
\caption{QSL and magnetic field in the magnetic flux rope (a) for the JL stable case at the time $t/t_0=85$, (b) for the JL unstable case at the time $t/t_0=110$. The gray-scale image on the bottom show the vertical magnetic field $B_z$. The magnetic field lines are colored by the magnetic field strength $|\mathbf{B}|$.
} \label{fig3}
\end{center}
(Two movies attached to this figure are available online.)
\end{figure}

\begin{figure}
\begin{center}
\includegraphics[width=0.8\textwidth]{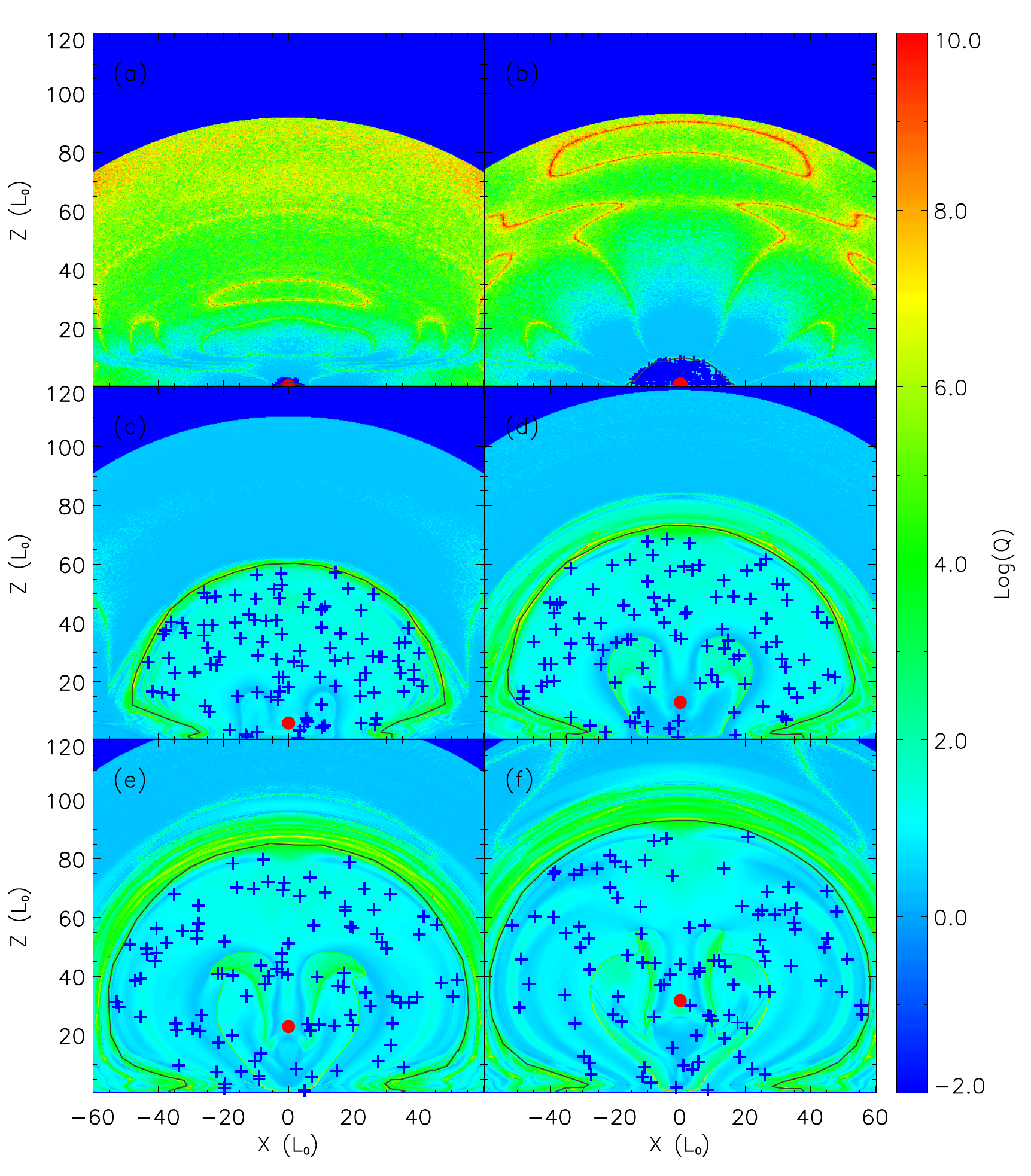}
\caption{Distribution of the squashing degree $Q$ in the $xz$-plane at $y=0$, namely, in the middle of the flux rope along its axis, for the JL stable models with $t/t_0=$ (a) 30, (b) 50, (c) 85, (d) 120, (e) 155, and (f) 190. Grey solid line delineates the boundary of the flux rope. A red dot indicates the position of the axis. Blue plus signs indicate the intersection of the sample field lines, used to retrieve the flux rope twist, with the $xz$-plane.
} \label{fig4}
\end{center}
\end{figure}

Figures~\ref{fig3}(a) and (b) show the QSLs and some selected magnetic field lines for the JL stable case at the time $t/t_0 = 85$ and for the JL unstable case at the time $t/t_0 = 110$. The involved basic configuration for both cases consist of highly sheared and twisted field lines surrounded by a prominent QSL. All field lines within the QSL are regarded as a coherent magnetic flux rope. The slice for the computation of $Q$ is placed within the $xz$-plane, centered at the middle point between all of the footpoints of the flux rope. Figure~\ref{fig4} displays the evolution of $Q$ within the same slice as a function of time for the JL stable case at times $t/t_0 = 30, 50, 85, 120, 155, 190$. Evidently, the QSL rises and expands over time. Similar to the Titov--D\'emoulin model, the axis of the magnetic flux rope is determined as the magnetic field line that is oriented most perpendicular with respect to the vertical slice. As shown in Figure~\ref{fig4}, the axis of the magnetic flux rope also rises with time. Table~\ref{tbl2} lists the writhe of each axis curve computed with Equations~(\ref{eqn:polar}), (\ref{eqn:local}), and (\ref{eqn:nonlocal}). Some field lines are selected randomly for the computation of the twist numbers within the QSL, their cross sections with the $xz$-plane indicated by the blue plus signs in Figure~\ref{fig4}.

\begin{figure}
\begin{center}
\includegraphics[width=0.7\textwidth]{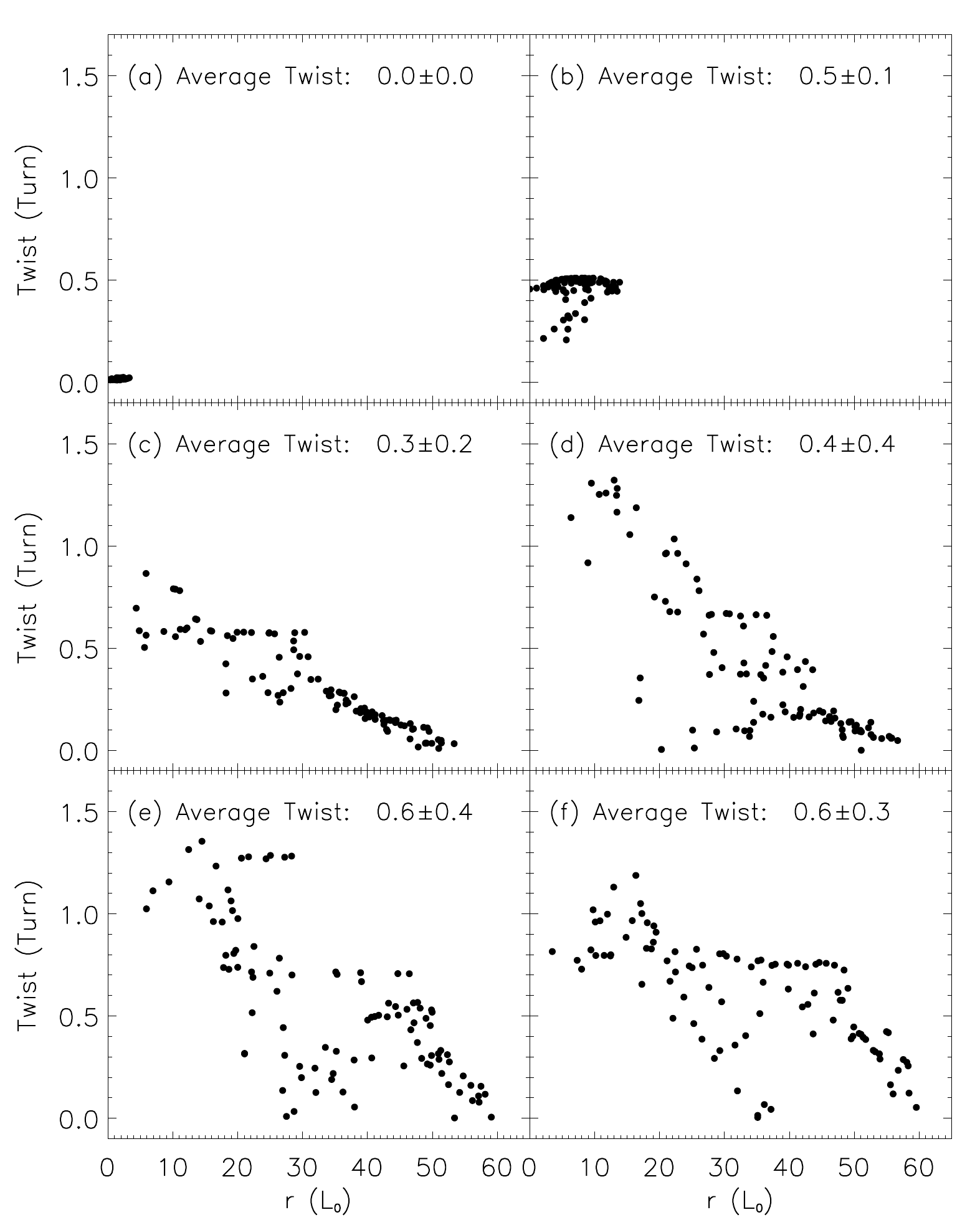}
\caption{Twist of the sample magnetic field lines along the distance, $r$, which is measured in the $xz$-plane at $y=0$, for the JL stable models with $t/t_0=$ (a) 30, (b) 50, (c) 85, (d) 120, (e) 155, and (f) 190.
} \label{fig5}
\end{center}
\end{figure}

Figure~\ref{fig5} displays the distributions of the twist numbers for the sample field lines as a function of distance to the axis curve of the flux rope, for the same time instances as before. In general, the field line closer to the axis possesses a larger twist. The average twist for each snapshot is indicated in Figure~\ref{fig5} and listed in Table~\ref{tbl2}. The magnetic fluxes within the QSL for each snapshot are also computed within the selected slices and listed in Table~\ref{tbl2}. Then, the magnetic helicity is computed by Equation~(\ref{eqn:hel2}) and listed in Table~\ref{tbl2}. The total relative magnetic helicity $\mathscr{H}_{V}$ has been computed in \citet{2016Valori} with six different volume helicity implementations, revealing a very small spread in the obtained helicity values. Therefore, in the following, we use the results obtained with the method of \citet{2012Valori}, based on the DeVore gauge, here for comparison. Together with the total relative magnetic helicity $\mathscr{H}_{V}$, the purely current-carrying part $\mathscr{H}_{V,J}$ are also computed with the DeVore gauge and listed in Table~\ref{tbl2}. A careful comparison shows that $\mathscr{H}_\mathrm{twist}$ matches $\mathscr{H}_{V,J}$ within the uncertainties for most snapshots except at $t/t_0 = 50$ and 85. The means of $\mathscr{H}_\mathrm{twist}/\mathscr{H}_{V,J}$ and $\mathscr{H}_\mathrm{twist}/\mathscr{H}_V$ for the JL stable cases (except the case JL-S-T30 where $\mathscr{H}_{V,J}$ and $\mathscr{H}_V$ are zero) are 1.99 and 0.16, respectively. The case JL-S-T85 contributes a large ratio of 5.88 for $\mathscr{H}_\mathrm{twist}/\mathscr{H}_{V,J}$. Excluding this case, the mean of $\mathscr{H}_\mathrm{twist}/\mathscr{H}_{V,J}$ for the other cases are 1.01. To compare the writhe and twist, we compute the mean of $|\mathcal{W}_p/\mathcal{T}|$, which is 0.089 for all cases except the JL-S-T30 case. The flux rope has barely emerged for the JL-S-T30 case and only a small portion is above the $z=0$ boundary, which might introduces a relatively large error in measuring the writhe and twist. We note that the uncertainties are quite large compared with the Titov--D\'emoulin cases in \citet{2016Valori}. In the JL cases, the structure of the flux rope is more extended than that in the Titov--D\'emoulin cases, resulting in the twist values varying significantly across the flux rope (hence, a large dispersion and standard deviation). The ratio $E_\mathrm{ns}/E$ is listed in the last column of Table~\ref{tbl2}, which shows that $E_\mathrm{ns}$ only contributes (at most) a few percent of the total magnetic energy. Based on the analysis in \citet{2016Valori}, we conclude that the error on helicity values due to the violation of the solenoidal property is small enough such that our conclusions are not affected by it.

\begin{figure}[h]
\begin{center}
\includegraphics[width=0.8\textwidth]{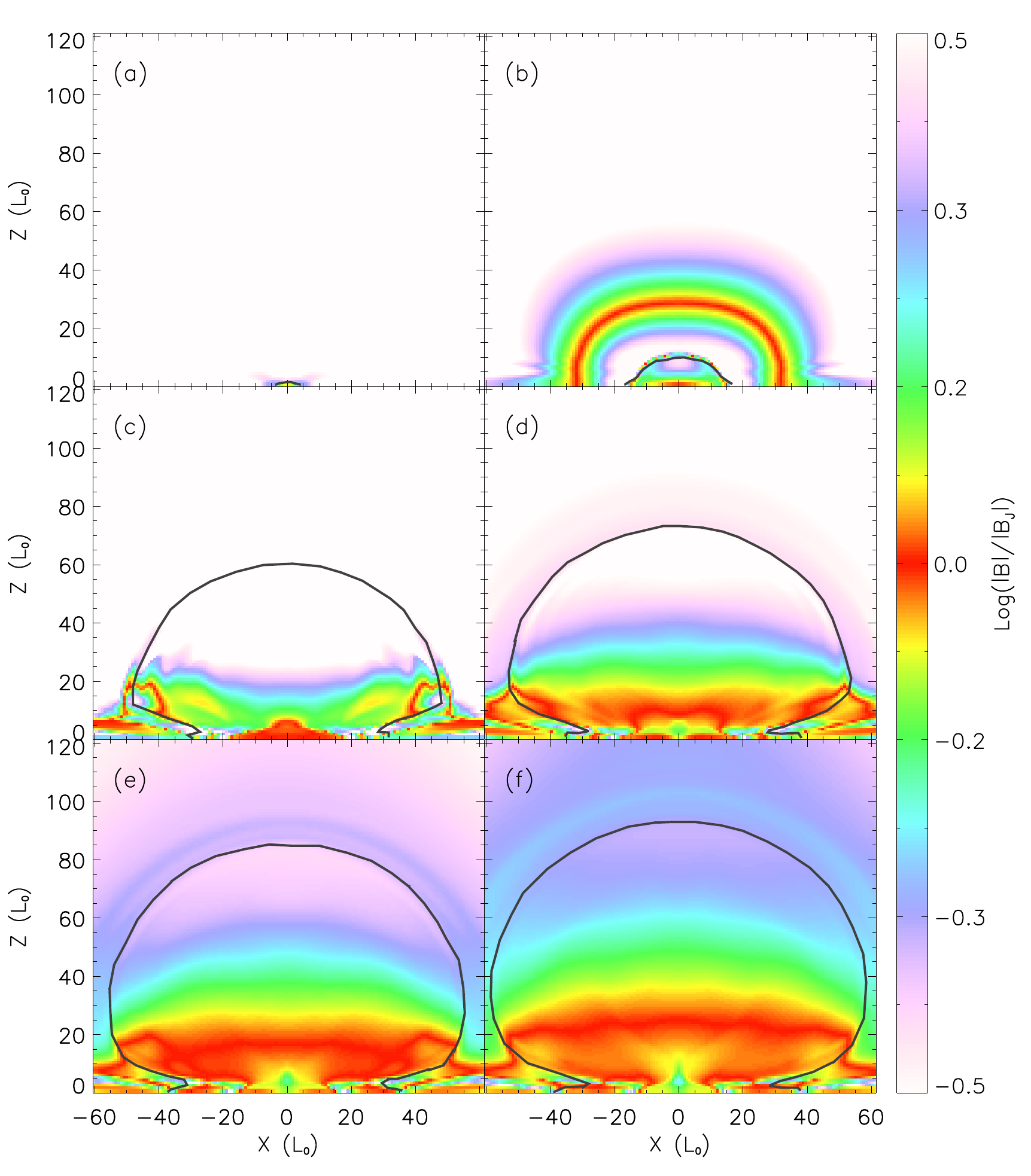}
\caption{Distribution of $\log(|\mathbf{B}|/|\mathbf{B}_J|)$, where $\mathbf{B}_J=\mathbf{B}-\mathbf{B}_\mathrm{p}$, for the JL stable case at $t/t_0=$ (a) 30, (b) 50, (c) 85, (d) 120, (e) 155, and (f) 190. Grey solid line, which is the same as that in Figure~\ref{fig4}, delineates the boundary of the flux rope.
} \label{fig6}
\end{center}
\end{figure}

Following Equations~(\ref{eqn:hv2}), (\ref{eqn:hvj}) and (\ref{eqn:hvjp}), the magnetic field helicity can be decomposed into the purely current-carrying part, and another part contributed by the potential field and the field generated by local currents. The above results show that the magnetic helicity, $\mathscr{H}_\mathrm{twist}$, computed by the twist method favorably compares with $\mathscr{H}_{V,J}$, the magnetic helicity purely contributed by the current-carrying part. To study the reason why $\mathscr{H}_\mathrm{twist}$ and $\mathscr{H}_{V,J}$ coincides within a magnetic flux rope, we decompose the magnetic field $\mathbf{B}$ into a potential part, $\mathbf{B}_\mathrm{p}$, and a current-carrying part $\mathbf{B}_J$, with $\mathbf{B} = \mathbf{B}_\mathrm{p} + \mathbf{B}_J$. Figure~\ref{fig6} displays the distribution of $\log(|\mathbf{B}|/|\mathbf{B}_J|)$ on the same slice as that for the squashing degree $Q$. Comparing with the $Q$ map in Figure~\ref{fig4}, we find that the regions where $\mathbf{B}_J$ contributes a major part of the total magnetic field, namely, $\log(|\mathbf{B}|/|\mathbf{B}_J|)$ is around 0, are mainly located within the QSL surrounding the magnetic flux rope. Therefore, the magnetic helicity computed by the twist method in the magnetic flux rope agrees with the helicity purely contributed by the current-carrying part for most snapshots. The result shown in Figure~\ref{fig6}(b) is an exception. The region where $\log(|\mathbf{B}|/|\mathbf{B}_J|)$ around 0 is outside of the QSL in Figure~\ref{fig4}(b). This is because the QSLs generally do not strictly correspond to the boundary of $\mathbf{B}_J$.

\begin{figure}
\begin{center}
\includegraphics[width=0.8\textwidth]{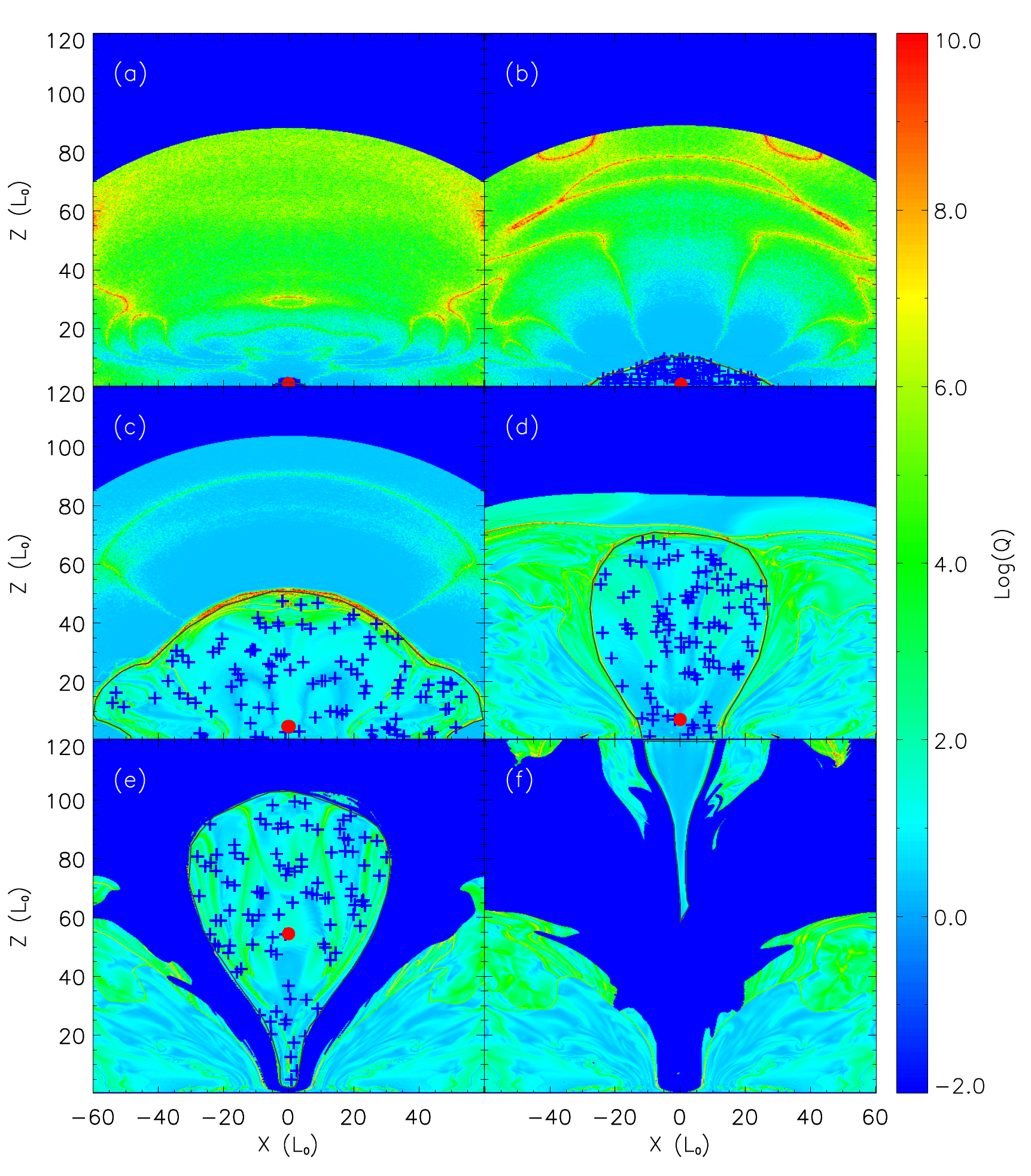}
\caption{Distribution of the squashing degree $Q$ in the $xz$-plane at $y=0$, namely, in the middle of the flux rope along its axis, for the JL unstable models with $t/t_0=$ (a) $30$, (b) $50$, (c) $80$, (d) $110$, (e) $140$, and (f) $150$. Grey solid line delineates the boundary of the flux rope. A red dot indicates the position of the axis. Blue plus signs indicate the intersection of the sample field lines, used to retrieve the flux rope twist, with the $xz$-plane.
} \label{fig7}
\end{center}
\end{figure}

We also compute the magnetic helicity with the twist number method for the JL unstable case. The squashing degree $Q$ is computed on a slice as shown in Figure~\ref{fig3}(b). The evolution of the $Q$ map at snapshots $t/t_0=30, 50, 80, 110, 140$ and $150$ is shown in Figure~\ref{fig7}. A magnetic flux rope is defined within the QSL delineated by large $Q$ values. The magnetic flux rope first rises (Figures~\ref{fig7}(a)--(d)), then detaches from the bottom boundary (Figure~\ref{fig7}(e)), and finally moves out of the selected area (Figures~\ref{fig7}(f)). The axis of the magnetic flux rope is determined by the magnetic field line that is orient most perpendicular with respect to the vertical slice along the $xz$-plane. The writhes of the axes at different times are listed in Table~\ref{tbl3}. Some sample field lines are randomly selected within the QSL surrounding the magnetic flux rope. In Figure~\ref{fig7}(f), we find that the axis of the magnetic flux rope has been propelled outside of the region of interest. Therefore, we do not compute the magnetic helicity for this snapshot.

\begin{figure}
\begin{center}
\includegraphics[width=0.7\textwidth]{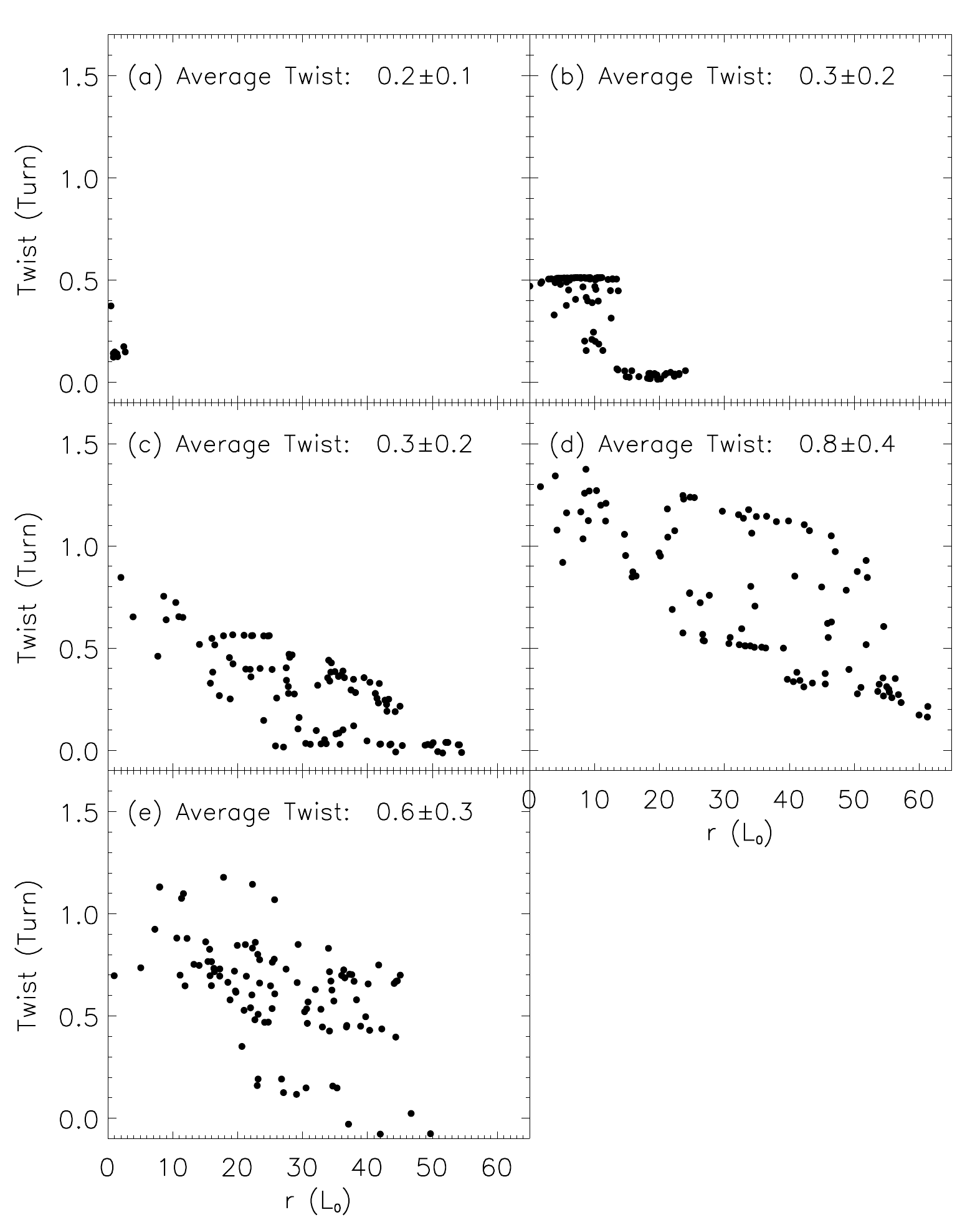}
\caption{Twist of the sample magnetic field lines along the distance, $r$, which is measured in the $xz$-plane at $y=0$, for the JL unstable models with $t/t_0=$ (a) $30$, (b) $50$, (c) $80$, (d) $110$, and (e) $140$.
} \label{fig8}
\end{center}
\end{figure}

The distributions of the twist numbers at $t/t_0 = 30, 50, 80, 110$ and $140$ along the distances to the axes are displayed in Figure~\ref{fig8}. Similar to the JL stable case, the twist also decreases from the axis to a distance further away from it. The average of the twist numbers is displayed in Figure~\ref{fig8} and listed in Table~\ref{tbl3}, where the magnetic flux and the magnetic helicities computed by the twist number method and finite volume method are also listed. The results show that $\mathscr{H}_\mathrm{twist}$ matches $\mathscr{H}_{V,J}$ within the uncertainties at $t/t_0=30, 50$, and 80, while $\mathscr{H}_\mathrm{twist}$ is less than $\mathscr{H}_{V,J}$ at $t/t_0=110$ and $140$. The means of $\mathscr{H}_\mathrm{twist}/\mathscr{H}_{V,J}$ and $\mathscr{H}_\mathrm{twist}/\mathscr{H}_V$ for the JL unstable cases (except the case JL-U-T30 where $\mathscr{H}_{V,J}$ and $\mathscr{H}_V$ are zero) are 0.58 and 0.35, respectively. The mean of $|\mathcal{W}_p/\mathcal{T}|$ is 0.091 for all the JL unstable cases except the JL-U-T30 case at $t/t_0=30$, at which time the writhe is large compared to its twist. We check its axis curve and find that it is highly coiled. However, it must be noted that $t=30t_0$ is a very early time in the emergence evolution referring to, e.g., Figure 2 in \citet{2014Leake}, where the proper identification of the axis might be prone to larger fluctuations. The ratio $E_\mathrm{ns}/E$ listed in the last column of Table~\ref{tbl3} indicates that the magnetic energy produced by the non-solenoidal part only contributes (at most) a few percent of the total magnetic energy.

\begin{figure}[h]
\begin{center}
\includegraphics[width=0.8\textwidth]{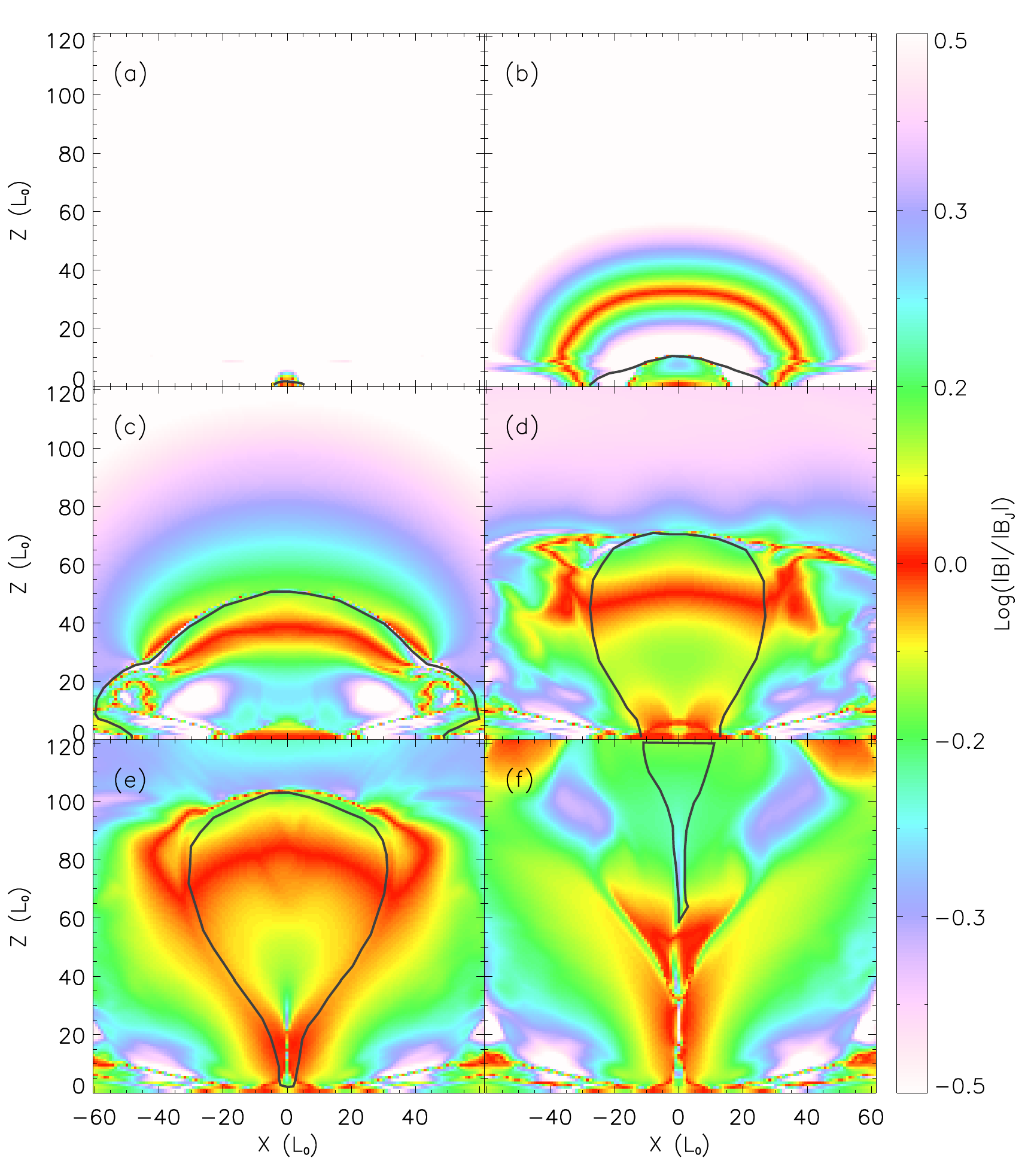}
\caption{Distribution of $\log(|\mathbf{B}|/|\mathbf{B}_J|)$, where $\mathbf{B}_J=\mathbf{B}-\mathbf{B}_\mathrm{p}$, for the JL unstable case at $t/t_0=$ (a) $30$, (b) $50$, (c) $80$, (d) $110$, (e) $140$, and (f) $150$. Grey solid line, which is the same as that in Figure~\ref{fig7}, delineates the boundary of the flux rope.} \label{fig9}
\end{center}
\end{figure}

To study the relationship between $\mathscr{H}_\mathrm{twist}$ and $\mathscr{H}_{V,J}$, we also decompose the magnetic field $\mathbf{B}$ of JL unstable model into a potential part, $\mathbf{B}_\mathrm{p}$, and a current-carrying part $\mathbf{B}_J$. Figure~\ref{fig9} displays the distribution of $\log(|\mathbf{B}|/|\mathbf{B}_J|)$ on the same slice (Figure~\ref{fig3}(b)) as that for the squashing degree $Q$. It is found that $\mathbf{B}_J$ contributes the major part of $\mathbf{B}$ mainly at three different places, namely, the front of the magnetic flux rope, the bottom boundary, and the current sheet stretched by the erupting magnetic flux rope. This point is different from the JL stable model, where $\mathbf{B}_J$ only contributes the major part close to the bottom boundary and within the magnetic flux rope (Figure~\ref{fig6}). This difference also explains why $\mathscr{H}_\mathrm{twist}$ within the magnetic flux rope deviates from $\mathscr{H}_{V,J}$ in the whole computation box. This is because there is large $\mathbf{B}_J$ outside of the magnetic flux rope as shown in Figures~\ref{fig9}(d) and (e). Figure~\ref{fig9}(b) is also an exception. The region where $\log(|\mathbf{B}|/|\mathbf{B}_J|)$ around 0 is outside of the QSL in Figure~\ref{fig7}(b).

\subsection{Nonlinear Force-Free Field Models} \label{sec:nlfff}

In this section, we compare the results obtained with the twist number method with that obtained with the aforementioned DeVore volume helicity method, using NLFFF extrapolations. In particular, we use two of the active regions studied in \cite{2015Savcheva,2016Savcheva}, one on 2007 February 12 and the other on 2010 August 7. The flux rope insertion method \citep{2004VanBallegooijen} is used to produce the NLFFF models. More details of this method are provided in \cite{2011Su} and \cite{2012Savcheva1}. The flux rope insertion method produces models that are in a wedge-shaped volume and in the spherical coordinate system. However, the helicity computation is performed in Cartesian coordinates, so we transform the spherical to Cartesian coordinates of the grid as in \cite{2012Savcheva2}.

\begin{figure}
\begin{center}
\includegraphics[width=0.8\textwidth]{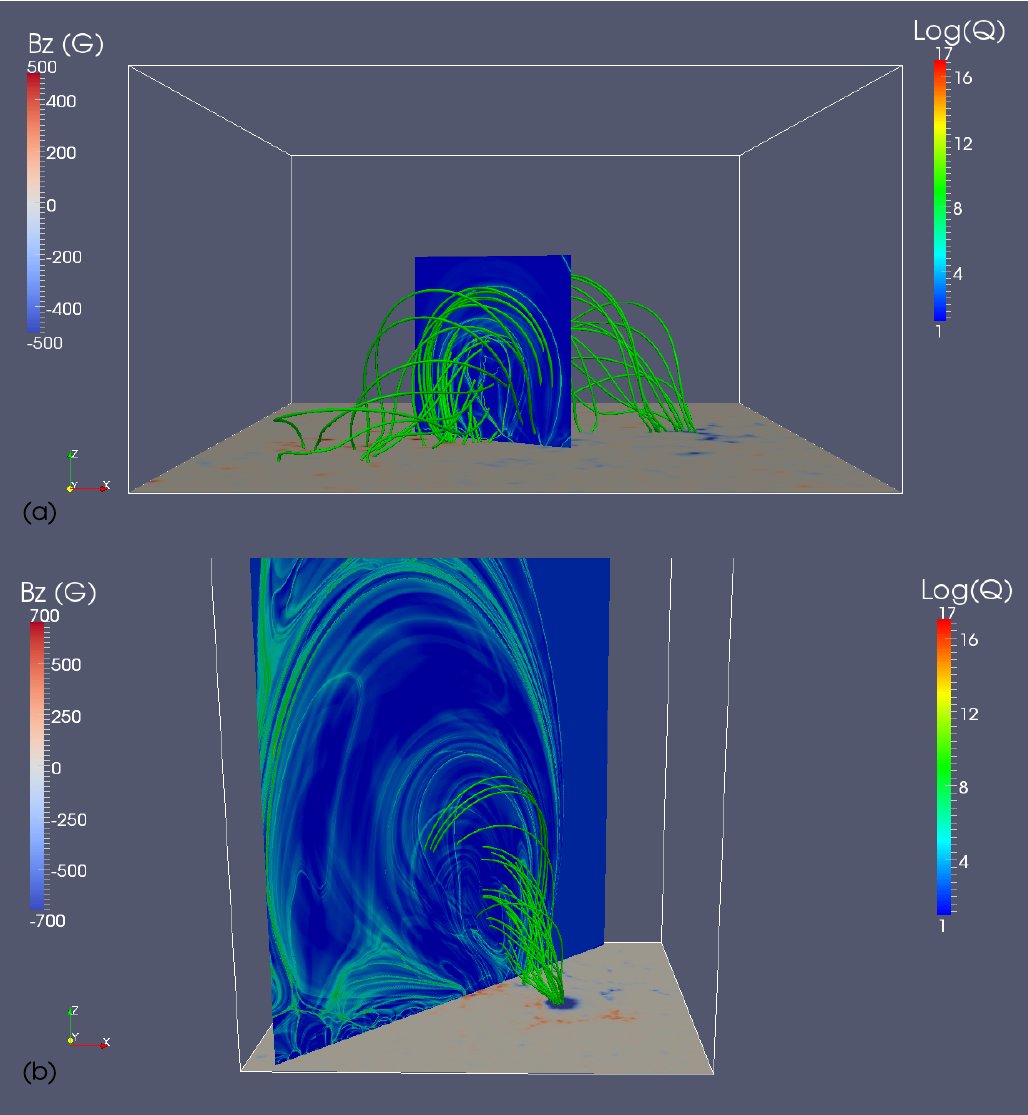}
\caption{QSL and magnetic field lines in the magnetic flux rope for the magnetic extrapolation models (a) at 06:41 UT on 2007 February 12 (b) at 17:00 UT on 2010 August 7. The color-scale image on the bottom show the vertical magnetic field $B_z$. The magnetic field lines are colored by a green color.
} \label{fig10}
\end{center}
(Two movies attached to this figure are available online.)
\end{figure}

The active region on 2007 February 12 produced a flare with a GOES class smaller than B starting at 07:40~UT. More details on the observations can be found in \citet{2015Savcheva}. Some selected magnetic field lines and QSLs on a slice of the NLFFF model are displayed in Figure~\ref{fig10}(a). The computation box in the range $[199.3, 420.4] \times [-256.4, -42.1] \times [0.0, 112.3]$ Mm is resolved into a uniform Cartesian grid of $227\times 220 \times 126$ pixels, with $(0, 0, 0)$ the coordinates of the central point on the disk. Since the axis of the magnetic flux rope is not along the $x$- or $y$-axis, the slice for the computation of $Q$ is selected at a oblique direction as shown in Figure~\ref{fig10}(a). We assume that the magnetic flux-rope axis is horizontal at the position where we cut the slice. And the normal direction of the slice points along the overall orientation of the flux rope delineating the flux-rope axis.

\begin{figure}
\begin{center}
\includegraphics[width=0.9\textwidth]{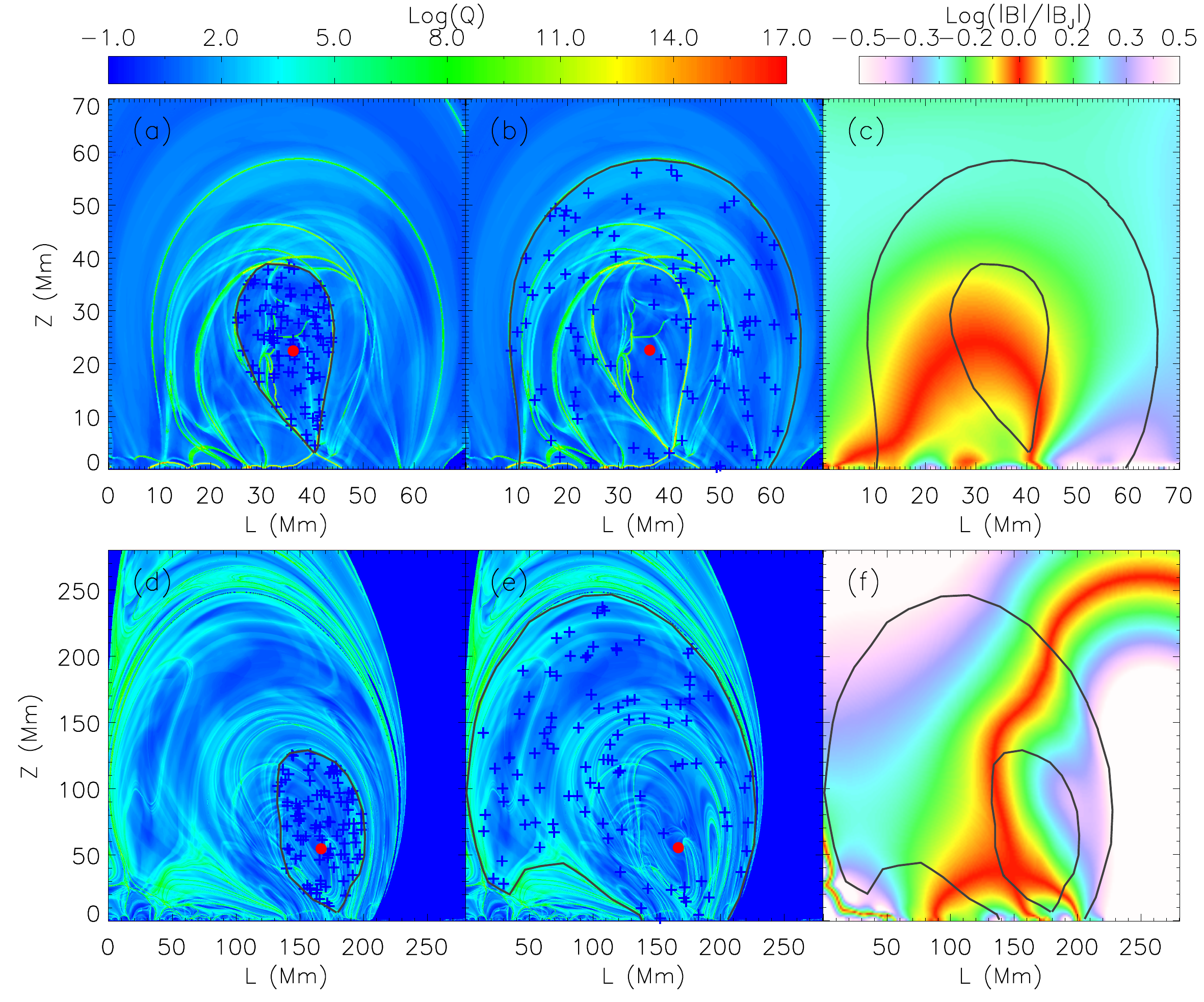}
\caption{Vertical slices of the $Q$ map that are perpendicular to the axes of the flux ropes, for the NLFFF models at (a, b) 06:41 UT on 2007 February 12,  and (d, e) 17:00 UT on 2010 August 7. Grey solid line delineates the boundary of the flux rope. Red dot indicates the position of the axis. Blue plus sign indicates the starting points of the sample field lines. Distribution of $\log(|\mathbf{B}|/|\mathbf{B}_J|)$, where $\mathbf{B}_J=\mathbf{B}-\mathbf{B}_\mathrm{p}$, for the slices at (c) 06:41 UT on 2007 February 12, and (f) 17:00 UT on 2010 August 7.
} \label{fig11}
\end{center}
\end{figure}

The axis of the magnetic flux rope is determined similarly to the previous cases, namely, it is the field line oriented most perpendicular with respect to the selected slice. This choice is supported by the resulting small writhe, as listed in Table~\ref{tbl4}. The red dot in Figures~\ref{fig11}(a) and (b) indicates the intersection of the axis of the magnetic flux rope with the vertical slice. However, there are some ambiguities in determining the body of the magnetic flux rope, because there are many interlaced QSLs on the cross section as shown in Figures~\ref{fig11}(a) and (b). This seems common for models constructed by NLFFF models \citep[e.g.,][]{2012Savcheva2,2012Savcheva1,2013Guo} since the magnetic field in observations is more complicated than in analytic and MHD models. The observed magnetic fields are usually distributed intermittently, which would introduce many bald patches \citep{2012Savcheva2} or magnetic null points \citep[e.g.,][]{2002Schrijver}. For comparison, we select two connectivity domains to define the body of the magnetic flux rope, one in an inner region surrounded by the QSL (Figure~\ref{fig11}(a)) and the other in a larger region surrounded by the outermost QSL (Figure~\ref{fig11}(b)). Some sample field lines, their intersection with the vertical slice denoted by blue crosses, are randomly selected in the two regions, respectively.

\begin{figure}
\begin{center}
\includegraphics[width=0.8\textwidth]{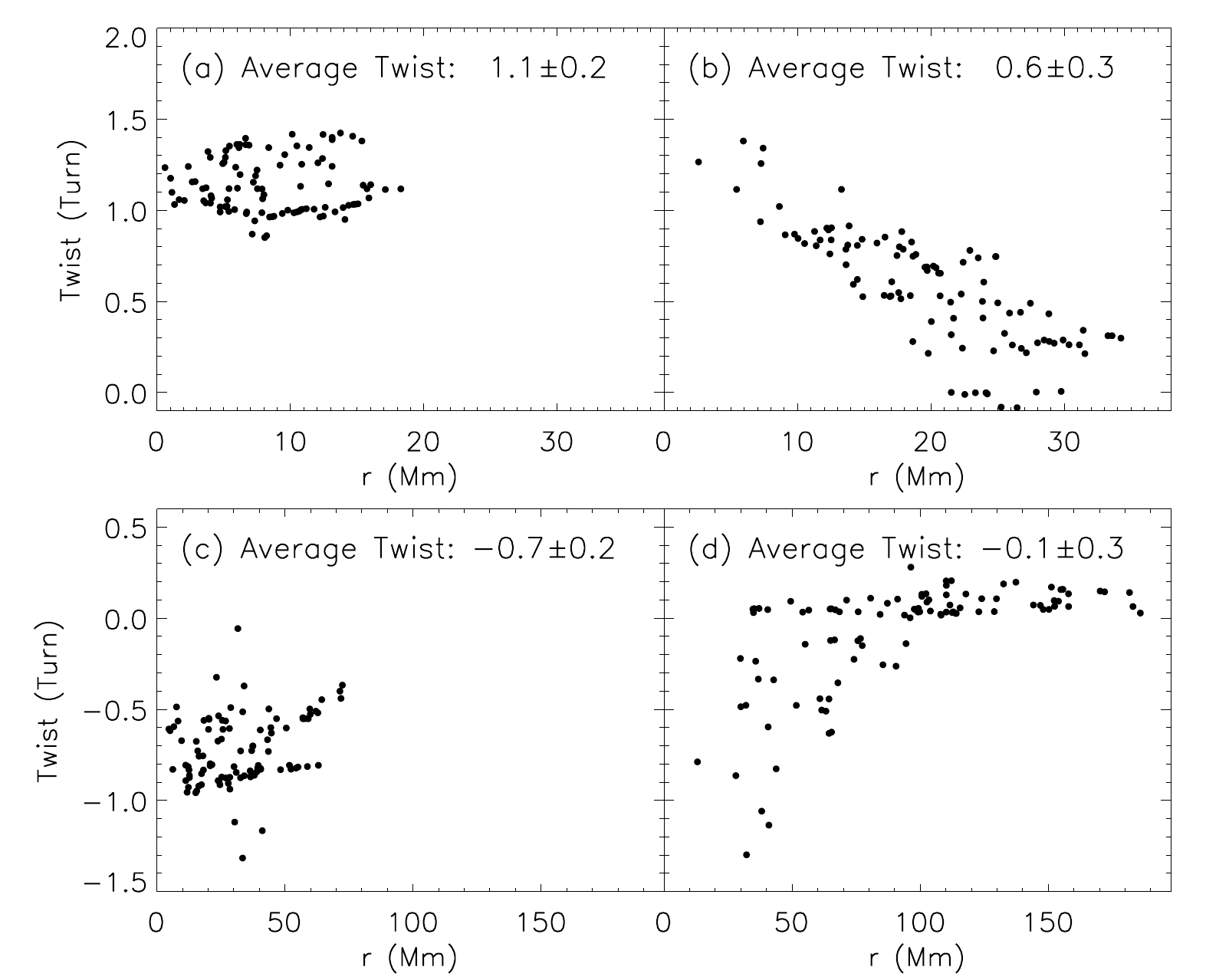}
\caption{Twist of the sample magnetic field lines along the distance, $r$ from the flux-rope axis, which is measured in the $Q$ slice plane as shown in Figure~\ref{fig10}, for the NLFFF models on (a) 2007 February 12 and the sample field lines in a small area as shown in Figure~\ref{fig11}(a), (b) 2007 February 12 and the sample field lines in a large area as shown in Figure~\ref{fig11}(b), (c) 2010 August 7 and the sample field lines in a small area as shown in Figure~\ref{fig11}(d), and (e) 2010 August 7 and the sample field lines in a large area as shown in Figure~\ref{fig11}(e).
} \label{fig12}
\end{center}
\end{figure}

The twist numbers of the sample field lines for the two different regions are displayed in Figures~\ref{fig12}(a) and (b), as a function of their distances of the sample field lines to the axis. The averages of the twist numbers are also shown in Figures~\ref{fig12}(a) and (b) and listed in Table~\ref{tbl4} as the smaller region case (NLFFF-S-20070212) and larger region case (NLFFF-L-20070212), respectively. Combined with the magnetic fluxes, the magnetic helicity, $\mathscr{H}_\mathrm{twist}$, can be computed by Equation~(\ref{eqn:hel2}). We find that $\mathscr{H}_\mathrm{twist}$ in the NLFFF-L-20070212 case matches $\mathscr{H}_{V,J}$ within the uncertainties.

Similar to the above analysis, some selected field lines and a $Q$ map for the sigmoidal active region on 2010 August 7 are shown in Figure~\ref{fig10}(b). Figures~\ref{fig11}(d) and (e) show two vertical slices, with different domains selected to outline the body of the flux rope. The distributions and averages of the twist numbers are displayed in Figures~\ref{fig12}(c) and (d). We list the writhe, twist, magnetic flux, and magnetic helicity (computed by the twist method and finite volume method) for both the smaller region (NLFFF-S-20100807) and the larger region (NLFFF-L-20100807) in Table~\ref{tbl4}. It is found that $\mathscr{H}_\mathrm{twist}$ is close to $\mathscr{H}_{V,J}$ for the NLFFF-S-20100807 case. The absolute values of the twist and magnetic flux for the NLFFF-L-20100807 case become even smaller than the NLFFF-S-20100807 case, which derives a small absolute value of $\mathscr{H}_\mathrm{twist}$. This is because the cross section in the NLFFF-L-20100807 case is too large. Many field lines far away from the axis become potential, which yields small twist numbers. At the same time, some magnetic field lines crossing the selected slice reverse their directions. This effect cancels the magnetic flux on the selected slice. Table~\ref{tbl4} also lists the ratio between the magnetic energy contributed by the non-solenoidal field and the total magnetic energy. This ratio is below a few percent.

The distributions of $\log(|\mathbf{B}|/|\mathbf{B}_J|)$ for the cases on 2007 February 12 and 2010 August 7 are shown in Figures~\ref{fig11}(c) and (f), respectively. Comparing Figures~\ref{fig11}(a) and (b) with Figure~\ref{fig11}(c), we find that $\mathbf{B}_J$ contributes the major part within the larger region (NLFFF-L-20070212) surrounded by the QSL. It explains why $\mathscr{H}_\mathrm{twist}$ of the NLFFF-L-20070212 case equals $\mathscr{H}_{V,J}$. Comparing Figures~\ref{fig11}(d) and (e) with Figure~\ref{fig11}(f), we also find that $\mathbf{B}_J$ contributes the major part within the larger region (NLFFF-L-200100807) surrounded by the QSL. But $\mathscr{H}_\mathrm{twist}$ is close to $\mathscr{H}_{V,J}$ in the smaller region (NLFFF-S-200100807). We have found that this is because the considered region is too large and the twist and magnetic flux for the flux rope are not accurate. Table~\ref{tbl4} also shows that the absolute value of $\mathscr{H}_\mathrm{twist}$ in the smaller region (NLFFF-S-20100807) is less than the absolute value of $\mathscr{H}_{V,J}$. This point can also be explained by the distribution of $\log(|\mathbf{B}|/|\mathbf{B}_J|)$ as shown in Figure~\ref{fig11}(f). The dominant region of $\mathbf{B}_J$, where $\log(|\mathbf{B}|/|\mathbf{B}_J|)$ is close to 0, is larger than the smaller region as shown in Figure~\ref{fig11}(d).

\section{Summary and Discussion} \label{sec:discussion}
We present a first systematic analysis of the magnetic helicity computed using the twist number method \citep{2010Guo,2013Guo}, and relate its performance to an existing and well-tested finite volume method. The input magnetic field configurations are either in the form of analytical (i.e., Titov--D\'emoulin) or numerical (MHD and NLFFF) models. The magnetic field models can be force-free (Titov--D\'emoulin and NLFFF) or not (MHD simulations). Our results suggest that the twist number method delivers helicity estimates, $\mathscr{H}_\mathrm{twist}$, in line with $\mathscr{H}_{V,J}$ (the magnetic helicity purely contributed by the current-carrying part) derived using a finite volume method, within the uncertainties for the semi-analytic models and MHD simulation cases. It also delivers similar values for NLFFF cases, given that the flux rope boundary is selected carefully.

To provide a quantitative comparison of $\mathscr{H}_\mathrm{twist}$, $\mathscr{H}_{V,J}$, and $\mathscr{H}_V$, we make some statistics for the ratio of the following three pairs of variables, namely, $\mathscr{\epsilon}_\mathrm{twist}/\mathscr{H}_\mathrm{twist}$ (where $\mathscr{\epsilon}_\mathrm{twist}$ marks the error of $\mathscr{H}_\mathrm{twist}$), $\mathscr{H}_\mathrm{twist}/\mathscr{H}_{V,J}$, and $\mathscr{H}_\mathrm{twist}/\mathscr{H}_V$. For the JL stable and unstable models, we exclude the cases JL-S-T30 and JL-U-T30, where all the values are zero and the ratios are not defined. For the NLFFF models, we only consider NLFFF-L-20070212 and NLFFF-S-20100807, where $\mathscr{H}_\mathrm{twist}$ is closer to $\mathscr{H}_{V,J}$ for each case than the other two cases. The means of $\mathscr{\epsilon}_\mathrm{twist}/\mathscr{H}_\mathrm{twist}$, $\mathscr{H}_\mathrm{twist}/\mathscr{H}_{V,J}$, and $\mathscr{H}_\mathrm{twist}/\mathscr{H}_V$ for all the considered cases are $51.1\%$, $125.6\%$, and $23.4\%$, respectively. We find that within the errors of $\mathscr{H}_\mathrm{twist}$, its value matches $\mathscr{H}_{V,J}$ better than $\mathscr{H}_V$. The agreement within error bars of $\mathscr{H}_\mathrm{twist}$ and $\mathscr{H}_{V,J}$ is valid for most models as demonstrated here and also in \citet{2016Valori}. The physical reason is that in a magnetic flux rope the magnetic field is mainly contributed by the local current. For those cases where $\mathscr{H}_\mathrm{twist}$ deviates from $\mathscr{H}_{V,J}$, the magnetic field is either fully dynamical (JL-U-T110 and JL-U-T140 as listed in Table~\ref{tbl3}) or topologically very complicated (NLFFF-S-20100807 and NLFFF-L-20100807 as listed in Table~\ref{tbl4}). In these cases, the local currents extend beyond the volume in the magnetic flux rope structure.

To quantify the contribution of the writhe to the self helicity in the test cases, we compute the polar writhe that is applicable to open curves \citep{2006Berger,2016Prior}. The mean of $|\mathcal{W}_p/\mathcal{T}|$ is 0.12 for all the cases listed in Tables~\ref{tbl2}, \ref{tbl3}, and \ref{tbl4} except the cases JL-S-T30, JL-U-T30, NLFFF-S-20070212, and NLFFF-L-20100807. Therefore, the magnetic helicity contributed by the twist is the dominant component over the part by the writhe. Together with the large uncertainties in computing the twist number, which is $51.1\%$ on average, the contribution of the writhe to the self helicity could be neglected in these test cases. For the cases with large writhe, the self helicity should be computed with $\mathcal{H}_\mathrm{self}=(\mathcal{W}_p+\mathcal{T})\Phi^2$, where both the writhe and twist are included. 

In terms of the fields topology, the magnetic helicity is divided into a self-helicity and mutual helicity as expressed in Equation~(\ref{eqn:hel}). Where the self-helicity is contributed by both the twist of a sample field line referred to its axis and the writhe of the axis itself. In this paper, we compute the helicity contributed by the twist; meanwhile, the helicity contributed by the writhe could be neglected, since most magnetic flux ropes do not have an obvious kink. Thus, $\mathscr{H}_\mathrm{twist}$ here is a good proxy for the self-helicity. While for the finite volume method as expressed in Equations~(\ref{eqn:hv2}), (\ref{eqn:hvj}), and (\ref{eqn:hvjp}), the magnetic helicity is divided by the purely current-carrying part, $\mathscr{H}_{V,J}$, and another part $\mathscr{H}_{V,JP}$. Since we have found that $\mathscr{H}_\mathrm{twist}$ (equivalent to the self-helicity in this paper) coincides with $\mathscr{H}_{V,J}$, it seems that $\mathscr{H}_{V,J}$ could also be interpreted as the self-helicity. We emphasize, however, that this interpretation is to be with caution, because the self-helicity is only a concept to help us to understand the magnetic helicity if one could regard the magnetic field to be composed by a finite number of flux tubes (or, flux ropes in case of an existing local current). As explained in \citet{2006Demoulin}, the self-helicity becomes negligible when the number of the flux tubes, $N$, approaches infinity, because the ratio between the self-helicity and the mutual helicity in Equation~(\ref{eqn:hel}) decreases as $1/N$.

Although the self-helicity ``vanishes'' when the magnetic field is thought of being composed of an infinite number of flux tubes, the concept is still useful. As demonstrated in this paper, we could derive the self-helicity for a very extended (with finite size) magnetic flux rope. And the results are comparable to those derived by the finite volume method. Under the assumption of a finite number of flux tubes, we might improve the computation with the aid of the mutual helicity method, e.g., the internal angle method proposed in \citet{2006Demoulin}. With this method, it is possible to divide an extended magnetic flux rope into smaller flux tubes, thus derive more accurately the magnetic helicity. Some preliminary tests have been presented in \citet{2016Yang}. Alternatively, one may use the method of the field line helicity \citep{2014Yeates,2016Yeates,2015Russell} to study the helicity flux distribution per field line. This method distinguishes the internal topology of a magnetic flux rope, and its integration over a cross section provides the total magnetic self-helicity. Corresponding in-depth studies are envisaged in the future.


\acknowledgments

This work is supported by the ISSI International Team on Magnetic Helicity estimations in models and observations of the solar magnetic field. The authors thank Bernhard Kliem, Tibor T\"or\"ok, James Leake, and Antonia Savcheva for making their numerical data available, and the referee for constructive comments on the computation of the writhe. YG is supported by NSFC (11533005 and 11203014), NKBRSF (2011CB811402 and 2014CB744203), the mobility grant from the Belgian Federal Science Policy Office (BELSPO), and the Poste Temporairement Vacant by the science council of the Observatoire de Paris. EP and KM acknowledge the support of the French Agence Nationale de la Recherche (ANR), under grant ANR-15-CE31-0001 (project HeliSol). GV acknowledges the support of the Leverhulme Trust Research Project Grant 2014-051. SA acknowledges support of the Russian Foundation of Basic Research under grants 150201089a, 150203835a, 16-32-00315-mol-a, and 153220504-mol-a-ved, and the Federal Agency for Scientific Organizations base project II.16.3.2 Non-stationary and wave processes in the solar atmosphere. KM and MKG were partially supported by the European Union (European Social Fund ESF) and Greek national funds through the Operational Program Education and Lifelong Learning of the National Strategic Reference Framework (NSRF) Research Funding Program: Thales. Investing in knowledge society through the European Social Fund. JKT acknowledges support from the Austrian Science Fund (FWF): P25383-N27. FC acknowledge the support by the International Max-Planck Research School (IMPRS) for Solar System Science at the University of G\"ottingen. SY is supported by grants KJCX2-EWT07 and XDB09040200 from the Chinese Academy of Sciences; grants 11221063, 11078012, 11178016, 11173033, 11125314, 10733020, 10921303, 11303053, 11573037, and 10673016 of National Natural Science Foundation of China; grant 2011CB811400 of National Basic Research Program of China; and KLSA2015 of the Collaborating Research Program of National Astronomical Observatories.



\clearpage

\clearpage

\clearpage

\clearpage

\clearpage

\clearpage

\clearpage

\clearpage

\clearpage

\clearpage

\clearpage

\clearpage

\clearpage

\clearpage

%

\clearpage

\begin{table}[h]
\caption{Writhe, Twist, and Magnetic Helicity of the JL Stable Models.} \label{tbl2}
\begin{tabular}{l r r r r r r r} \\
\hline \hline
Models     & Writhe      & Twist  &  Magnetic Flux     & $\mathscr{H}_\mathrm{twist}$   & $\mathscr{H}_V$   & $\mathscr{H}_{V,J}$   & $E_\mathrm{ns}/E$ \\
           & (Turn) & (Turn) & ($B_0 L_0^2$)      & ($10^{3} B_0^2 L_0^4$)  & ($10^{3} B_0^2 L_0^4$) & ($10^{3} B_0^2 L_0^4$)  &  \\
\hline
JL-S-T30   & 0.0033 & $0.02 \pm 0.00$   & $0.0$& $0.0 \pm 0.0$ & 0.00 & 0.00 & 0.058\\
JL-S-T50   & -0.013  & $0.47 \pm 0.06$  & $22$ & $0.2 \pm 0.0$ & 1.74 & 0.13 & 0.017\\
JL-S-T85   & 0.029   & $0.32 \pm 0.22$  & $55$ & $1.0 \pm 0.6$ & 6.98 & 0.17 & 0.011\\
JL-S-T120  & 0.067   & $0.42 \pm 0.38$  & $63$ & $1.6 \pm 1.5$ & 11.2 & 1.71 & 0.007\\
JL-S-T155  & 0.061   & $0.56 \pm 0.37$  & $72$ & $2.9 \pm 1.9$ & 14.3 & 3.41 & 0.005\\
JL-S-T190  & 0.037   & $0.61 \pm 0.27$  & $78$ & $3.6 \pm 1.6$ & 16.6 & 4.90 & 0.005\\
\hline
\end{tabular}
\end{table}

\begin{table}[h]
\caption{Writhe, Twist, and Magnetic Helicity of the JL Unstable Models.} \label{tbl3}
\begin{tabular}{l r r r r r r r} \\
\hline \hline
Models     & Writhe      & Twist  & Magnetic Flux  & $\mathscr{H}_\mathrm{twist}$       & $\mathscr{H}_V$   & $\mathscr{H}_{V,J}$   & $E_\mathrm{ns}/E$ \\
           & (Turn) & (Turn) & ($B_0 L_0^2$)  & ($10^{3} B_0^2 L_0^4$) & ($10^{3} B_0^2 L_0^4$) & ($10^{3} B_0^2 L_0^4$)  &  \\
\hline
JL-U-T30   & -0.63    & $0.16 \pm 0.07$  & $0.0$& $0.0 \pm 0.0$ & 0.00 & 0.00 & 0.058\\
JL-U-T50   & 0.00020   & $0.35 \pm 0.20$ & $25$ & $0.2 \pm 0.1$ & 0.40 & 0.25 & 0.011\\
JL-U-T80   & 0.043    & $0.29 \pm 0.21$  & $49$ & $0.7 \pm 0.5$ & 1.49 & 0.92 & 0.008\\
JL-U-T110  & 0.12     & $0.76 \pm 0.36$  & $35$ & $0.9 \pm 0.4$ & 2.77 & 1.67 & 0.005\\
JL-U-T140  & -0.035   & $0.62 \pm 0.26$  & $26$ & $0.4 \pm 0.2$ & 3.85 & 1.68 & 0.004\\
\hline
\end{tabular}
\end{table}

\clearpage

\begin{table}[h]
\caption{Writhe, Twist, and Magnetic Helicity of the NLFFF Models.} \label{tbl4}
\footnotesize
\begin{tabular}{r r r r r r r r} \\
\hline \hline
Models    & Writhe      & Twist   & Magnetic Flux     & $\mathscr{H}_\mathrm{twist}$     & $\mathscr{H}_V$   & $\mathscr{H}_{V,J}$   & $E_\mathrm{ns}/E$ \\
          & (Turn) & (Turn)  & ($10^{20}$Mx)   & ($10^{40}$ Mx$^2$) & ($10^{40}$ Mx$^2$) & ($10^{40}$ Mx$^2$)  & \\
\hline
NLFFF-S-20070212  & 0.22   & $1.14 \pm 0.15$   & $1.8$  & $3.6 \pm 0.5$    & $106 \pm 11$ & $20.5 \pm 0.2$ & 0.043\\
NLFFF-L-20070212  & 0.22   & $0.57 \pm 0.32$   & $5.8$  & $19.5 \pm 11.0$  & $106 \pm 11$ & $20.5 \pm 0.2$ & 0.043 \\
NLFFF-S-20100807  & -0.072 & $-0.72 \pm 0.19$  & $11.5$ & $-95.7 \pm 25.4$ & $-570 \pm 97$ & $-164 \pm 1$  & 0.061\\
NLFFF-L-20100807  & -0.072 & $-0.09 \pm 0.31$  & $3.3$  & $-0.9 \pm 3.3$   & $-570 \pm 97$ & $-164 \pm 1$  & 0.061\\
\hline
\end{tabular}
\end{table}

\end{document}